\documentclass[journal]{IEEEtran}
\usepackage{cite}
\usepackage{amsmath,amssymb,amsfonts,dsfont}
\usepackage[noend]{algpseudocode}
\usepackage{algorithmicx}
\usepackage{algorithm}
\usepackage{graphicx}
\usepackage{subfigure}
\usepackage{textcomp}
\usepackage{xcolor}
\usepackage{booktabs}
\usepackage{breqn}
\usepackage{subeqnarray}
\usepackage{cases}
\usepackage{setspace}
\usepackage{amsthm}
\usepackage{bbm}
\usepackage{bm}
\usepackage{lipsum}
\usepackage{enumitem}
\usepackage{mathtools}
\usepackage{bbm}
\usepackage{makecell}
\usepackage{multirow}
\usepackage{amsmath, amssymb}
\usepackage{booktabs}

\allowdisplaybreaks[4]

\theoremstyle{remark}

\newtheorem{rem}{Remark}

\def\BibTeX{{\rm B\kern-.05em{\sc i\kern-.025em b}\kern-.08em
		T\kern-.1667em\lower.7ex\hbox{E}\kern-.125emX}}

\begin{document}
	\title{Efficient LLM Inference over Heterogeneous Edge Networks with Speculative Decoding \\
	}
	\author{Bingjie~Zhu, Zhixiong~Chen,~\IEEEmembership{Member,~IEEE,}   Liqiang~Zhao,~\IEEEmembership{Member,~IEEE,}\\
		Hyundong~Shin,~\IEEEmembership{Fellow,~IEEE}, and Arumugam~Nallanathan,~\IEEEmembership{Fellow,~IEEE}
		\thanks{Bingjie Zhu is with the State Key Laboratory of Integrated Service Networks, Xidian University, Xi'an 710071, China. (e-mail: bjzhu1@stu.xidian.edu.cn).}
		\thanks{Zhixiong Chen and Arumugam Nallanathan are with the School of Electronic Engineering and Computer Science, Queen Mary University of London, London, U.K. Arumugam Nallanathan is 	also with the Department of Electronic Engineering, Kyung Hee University, Yongin-si, Gyeonggido 17104, Korea. (emails: \{zhixiong.chen, a.nallanathan\}@qmul.ac.uk).}
		\thanks{Liqiang Zhao is with the State Key Laboratory of Integrated Service Networks, Xidian University, Xi'an 710071, China, and also with Guangzhou Institute of Technology, Xidian 	University, Guangzhou 510100, China. (lqzhao@mail.xidian.edu.cn).}
		\thanks{Hyundong Shin is with the Department of Electronics and Information Convergence Engineering, Kyung Hee University, Yongin-si, Gyeonggido 17104, Republic of Korea (e-mail: hshin@khu.ac.kr).}
	}		
	
	\maketitle
	\begin{abstract}
	Large language model (LLM) inference at the network edge is a promising serving paradigm that leverages distributed edge resources to run inference near users and enhance privacy.
	Existing edge-based LLM inference systems typically adopt autoregressive decoding (AD), which only generates one token per forward pass.
	This iterative process, compounded by the limited computational resources of edge nodes, results in high serving latency and constrains the system’s ability to support multiple users under growing demands.
	To address these challenges, we propose a speculative decoding (SD)–based LLM serving framework that deploys small and large models across heterogeneous edge nodes to collaboratively deliver inference services. 
	Specifically, the small model rapidly generates draft tokens that the large model verifies in parallel, enabling multi-token generation per forward pass and thus reducing serving latency. 
	To improve resource utilization of edge nodes, we incorporate pipeline parallelism to overlap drafting and verification  across multiple inference tasks.
	Based on this framework,
	we analyze and derive a comprehensive latency model incorporating
	both communication and inference latency. Then, we
	formulate a joint optimization problem for speculation length,
	task batching, and wireless communication resource allocation to
	minimize total serving latency. To address this problem, we derive
	the closed-form solutions for wireless communication resource allocation, and develop a dynamic programming algorithm for joint
	batching and speculation control strategies.
	Experimental results demonstrate that the proposed framework achieves lower serving  latency compared to AD-based serving systems. 
	In addition, the proposed joint optimization method delivers up to 44.9\% latency reduction compared to benchmark schemes.	
	
	\end{abstract}
	
	\begin{IEEEkeywords}
		LLM, speculative decoding, batching, edge networks, wireless resource allocation.
	\end{IEEEkeywords}
	
	\section{Introduction}
	Recent advancements in large language models (LLMs), exemplified by ChatGPT, have revolutionized natural language understanding and generation capabilities \cite{qu2025mobile}.
	Open-source LLMs, such as the LLaMA family \cite{touvron2023llama}, further extend this potential by enabling task-specific fine-tuning and supporting diverse applications across healthcare, finance, and virtual assistance domains \cite{zhou2024large}.
	However, these advances are accompanied by increasingly large model sizes, resulting in substantial computational and memory requirements for LLM inference.
	Consequently, LLMs are currently deployed in centralized cloud centers with powerful computational capacity to provide services.
	Nevertheless, this centralized architecture necessitates transmitting user data to remote cloud servers for inference, thereby introducing significant transmission latency and raising critical data privacy concerns.
	These limitations may hinder the widespread application of LLMs in latency- and privacy-sensitive scenarios such as autonomous vehicles, real-time robotic control, and healthcare systems \cite{lin2025pushing}.

	Edge inference has emerged as a compelling paradigm that mitigates these issues by utilizing edge computing resources to provide inference services closer to users, which reduces transmission overhead and enhances privacy protection.
	Generally, edge inference can be categorized into single-node inference and multi-node collaborative inference.
	In single-node inference, the complete model is deployed on an edge server to process user requests using received task data and transmit inference results back to users via wireless links.
	Several studies have explored optimization techniques for single-node inference. 
	In \cite{cang2024joint}, the authors investigated batching and task scheduling techniques  to improve the completion rate of deep neural network (DNN) inference tasks in single-cell, multi-user scenarios.
	To maximize system throughput under limited resource constraints, Liu et al. \cite{liu2023resource} combined batching with early exiting, and jointly optimized communication and computation resources for multi-user edge inference.
	Despite advances such as early exiting and batching, a single resource-constrained edge server may struggle to support numerous concurrent inference tasks while maintaining low-latency serving. 
	To mitigate this issue, multi-node collaborative inference distributes models among edge nodes to complete inference tasks cooperatively based on the computing capacity of edge servers.
	In \cite{yao2022loading}, the authors investigated the joint optimization of model caching and request routing to maximize throughput in multi-edge server scenarios. 
	Li et al. \cite{li2024optimal} proposed a device-edge co-inference framework that jointly optimizes model partitioning and resource allocation to minimize energy consumption in wireless sensing systems.
	
	Although the aforementioned approaches in \cite{liu2023resource,cang2024joint,yao2022loading,li2024optimal} effectively enhance the performance of edge inference systems, they primarily focus on traditional DNNs.
	Traditional DNNs, such as MobileNet and ResNet, typically require only a single forward pass through the network with fixed computational requirements to generate inference results.
	In contrast, LLMs are characterized by larger model sizes  and reliance on autoregressive decoding (AD) for inference.
	This AD process generates inference results, i.e., output tokens, through iterative forward passes, with each forward pass producing one token conditioned on both the initial input prompt and all previously generated tokens.
	To accelerate this inherently sequential process, LLMs utilize key-value (KV) caches that store the hidden states of previous tokens, enabling only the newest token to be processed at each step while avoiding redundant computation.
	However, the KV cache grows linearly with the output sequence length, increasing memory consumption.
	Due to large model sizes, the AD mechanism, and KV cache, LLM inference often demands substantially greater computational and memory resources than traditional DNN inference.
	For example, a widely adopted DNN such as MobileNetV2, requires approximately 7 MB of memory and 0.3 GFLOPs for single-image inference whereas even a compact 3B-parameter LLM such as Phi-2 demands at least 6.3 GB of memory and 6 TFLOPs to generate an output sequence with 1000 tokens in 16-bit floating-point (FP16) precision.
	The significant disparity in resource requirements and inference mechanisms renders existing DNN-oriented edge inference approaches ineffective for LLM inference at the resource-limited edge. 

	Existing works  have explored approaches to enhance the efficiency of LLM inference over edge networks \cite{zhang2024beyond,11044591,zhang2024edgeshard, 11045182,xie2025mixture, he2024large}, typically targeting edge-deployable models up to 13B parameters, such as LLaMA-1.1B, LLaMA-7B, and OPT-13B. 
	These works mainly focus on model quantization \cite{zhang2024beyond,11044591}, model parallelism \cite{zhang2024edgeshard,11045182}, and resource management \cite{xie2025mixture,he2024large}.
	Specifically, the authors in \cite{zhang2024beyond} developed  model quantization approaches to reduce the computational costs of LLM inference and improve system throughput.
	In \cite{11044591}, the authors proposed an LLM scheduling framework that integrates model quantization with heterogeneous edge resource allocation to reduce operational costs.
	Pipeline parallelism was employed in \cite{zhang2024edgeshard} to reduce inference latency by partitioning model layers among edge nodes, enabling efficient overlapped execution of different computational stages.
	In \cite{11045182}, the authors employed tensor parallelism to distribute LLM tensors across edge nodes and used over-the-air computation to reduce communication overhead incurred by  all-reduce operations.
	The authors in \cite{xie2025mixture}  proposed a mixture-of-experts–enabled LLM inference framework, incorporating a Stackelberg game–based resource allocation mechanism to improve inference efficiency and resource utilization.
	An active inference-based algorithm was proposed in \cite{he2024large} to optimize task offloading and resource allocation policies for LLM inference tasks in cloud-edge collaborative architectures.
	
	Although the above approaches improve LLM inference efficiency at the edge, model lightweight techniques \cite{zhang2024beyond,11044591} may compromise output accuracy while model parallel methods \cite{zhang2024edgeshard, 11045182} suffer from frequent synchronization overhead between edge nodes.
	Furthermore, all methods in \cite{zhang2024beyond,11044591,zhang2024edgeshard, 11045182,xie2025mixture, he2024large} rely on AD to complete inference tasks, which is inherently constrained by sequential token generation. 
	This sequential pattern limits GPUs to single-token generation per forward pass of LLMs, underutilizing their parallel processing capabilities designed for simultaneous computation across thousands of cores.
	Consequently, this bottleneck may result in high serving latency, making efficient LLM serving system on resource-constrained edge devices challenging.
	To address the above issues, this work aims to develop an efficient distributed LLM inference system over edge networks that provide low-latency services without accuracy degradation.
	Specifically, we adopt speculative decoding (SD) for LLM inference over edge networks.
	SD employs a lightweight draft model to generate multiple speculative tokens, which a larger LLM (referred to as the target model) then verifies in parallel based on acceptance criteria \cite{leviathan2023fast,kim2023speculative}.
	This parallel verification approach effectively utilizing GPU parallel computing capabilities, and enables simultaneous generation of multiple tokens in a forward pass of the target model while maintaining target model accuracy \cite{spector2023accelerating,sadhukhan2024magicdec}.
	In addition, since the small draft model  incurs low computational costs, SD could efficiently accelerate LLM inference compared to AD.
	To leverage the advantages of SD, we deploy draft and target models across heterogeneous edge nodes.
	To improve resource utilization of edge nodes, we incorporate pipeline parallelism in the SD-based LLM inference.
	The pipeline parallelism overlaps draft speculation and target verification across different inference tasks, allowing the system to process multiple inference tasks concurrently and thus reducing total serving latency.
	The main contributions of this paper are summarized as follows:

	\begin{itemize}
		\item 
		We propose a novel SD-based LLM inference framework at the heterogeneous edge networks, in which draft and target models are deployed across different edge nodes to collaboratively provide LLM inference services. 
		This framework effectively reduces LLM inference latency while preserving model accuracy.
		
		\item We design a pipeline-parallel LLM serving mechanism with SD, which efficiently improves computational resource utilization across edge nodes.
		Based on this serving mechanism, we analyze and derive the total serving latency, incorporating both communication and inference latency.
		We then formulate a joint optimization problem for speculation length, batching, and wireless communication resource allocation to minimize total serving latency. This optimization problem is challenging to solve due to the tight coupling among decision variables in the latency formulation.

		\item To address this problem, we first derive the optimal wireless communication resource allocation policy for each inference task. Subsequently, we develop a dynamic programming-based algorithm to obtain efficient batching decisions and speculation control policies with low complexity.
		
		\item 
		We conduct extensive simulations to validate the effectiveness of the proposed LLM serving framework. 
		Specifically, compared to the AD-based serving systems, the proposed LLM serving system consistently delivers lower serving latency across three LLaMA model combinations ranging from 68M to 13B parameters.
		In addition, the proposed joint optimization policy achieves up to 44.9\%  latency reduction compared to benchmark schemes.

	\end{itemize}
	The rest of this paper is organized as follows: Section \ref{sec:system_model} introduces the system model, the proposed LLM serving system, and the problem formulation.
	Section \ref{sec:algor} illustrates the proposed speculation length, batch scheduling, and communication allocation algorithm.
	Section \ref{sec:simulation} evaluates the effectiveness of the proposed approaches by simulations.
	The conclusion is presented in Section \ref{sec:conclusion}.

	\section{System Model and Problem Formulation}\label{sec:system_model}
\begin{figure}[t]
		\centering
		\includegraphics[width=0.45\textwidth]{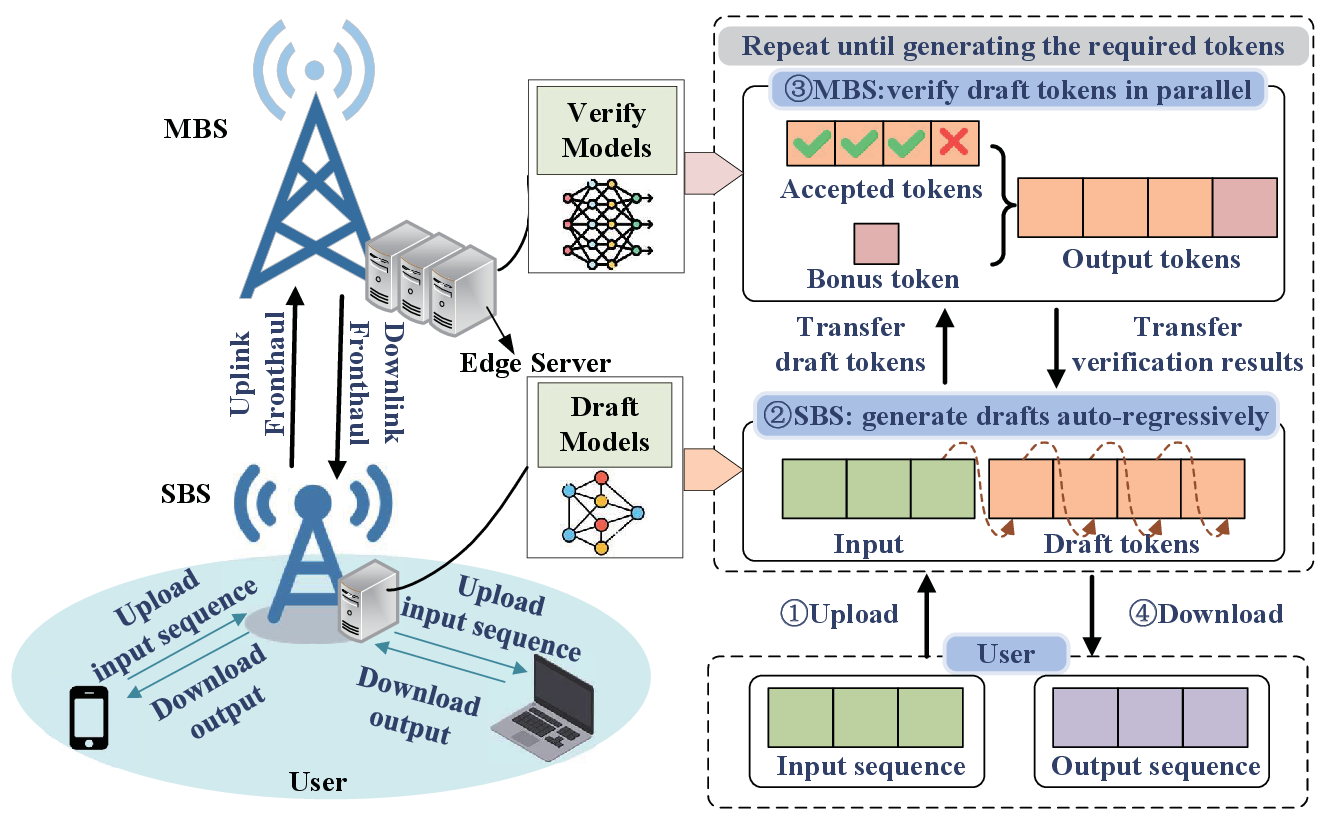}\\
		\caption{Illustration of the proposed LLM serving system with speculative decoding.}
		\label{fig:sys_model}
	\end{figure}

	This work investigates an LLM serving system in an edge network, which consists of a macro base station (MBS) and a small base station (SBS) to provide LLM services for multiple users, as shown in Fig. \ref{fig:sys_model}.
	The SBS is co-located with an edge server having certain computing and memory resources, while the MBS possesses strong computing and memory capabilities as it is usually fiber-optically connected to multiple edge servers \cite{8894168}.
	The MBS and the SBS are interconnected via a wired fronthaul link to 
	collaboratively serve users.
	The set of LLM inference tasks from users is denoted by $\mathcal{K}=\{1,2,...,K\}$, where each inference task $k$ is characterized by a 2-item tuple $\{I_{k},O_{k}\}$. 	
	Here, $I_k$ and $O_k$ denote the length of the input token sequence and the expected number of output tokens for inference task $k$, respectively.
	LLM inference typically adopts AD, generating one output token per forward pass given an input sequence. 
	However, AD incurs high inference latency, as generating a complete output token sequence requires multiple forward passes through the model.
	SD mitigates this issue by introducing a small  draft model to predict future output tokens.
	The target LLM (referred to as the verify model) then verifies which of those tokens to accept in parallel, thereby accelerating LLM inference.
	The verify model generally has substantial computational requirements, exceeding the processing capabilities of the SBS.
	For example, many existing works use LLaMA-7B as the verify model and LLaMA-68M as the draft model \cite{miao2024specinfer}. 
	LLaMA-7B requires approximately 14 GB of memory to load its model parameters in FP16 precision during inference, exceeding the capacity of most edge servers, such as the NVIDIA Jetson TX2 with only 8 GB of memory \cite{nvidia_jetson_tx2}. 
	Motivated by this, we deploy the verify model at the MBS with strong computational capacity, and the small draft model at the SBS.
	By executing LLM inference with SD, the SBS and MBS collaboratively 
	generate output tokens for each  task.
	
	\subsection{Inference Mechanism with SD} \label{subsec:inference_model}
		\begin{figure}[t]
		\centering
		\includegraphics[width=0.45\textwidth]{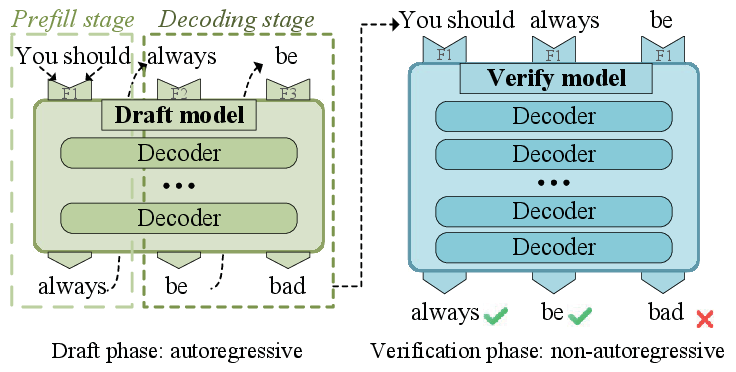}\\
		\caption{The typical workflow of speculative decoding.}
		\label{fig:SD}
	\end{figure}
	\begin{figure}[t]
		\centering
		\includegraphics[width=0.45\textwidth]{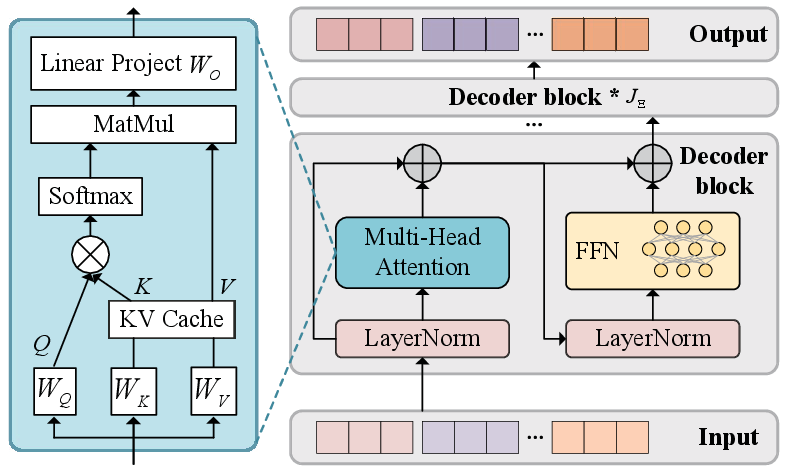}\\
		\caption{The computation procedure of transformer decoder-only models.}
		\label{fig:tranf}
		\vspace{-10pt} 
	\end{figure}
	To characterize the inference procedure, we first describe the structures of the draft model and the verify model. 
	Similar to many existing works~\cite{kim2023speculative,spector2023accelerating}, the two models adopt a decoder-only architecture composed of multiple stacked identical Transformer decoder blocks, as illustrated in Fig.~\ref{fig:SD}.
	For ease of presentation, we define a variable $\Xi \in\{\rm{v,d}\}$ to represent the model index, where $\rm{v}$ and $\rm{d}$ denote the verify model and the draft model, respectively.
	As shown in Fig. \ref{fig:tranf}, each decoder block in model $\Xi$ consists of a multi-head attention (MHA) module with weight matrices  $\mathbf{W}_Q, \mathbf{W}_K, \mathbf{W}_V, \mathbf{W}_O\in \mathcal{R}^{h_1^\Xi \times h_1^\Xi}$,
	followed by a fully connected feed-forward network (FFN) parameterized by $\mathbf{W}_1 \in \mathcal{R}^{h_1^\Xi \times h_2^\Xi}$ and $\mathbf{W}_2\in \mathcal{R}^{h_2^\Xi \times h_1^\Xi}$.
	Here, $h_1^\Xi$ and $h_2^\Xi$ represent the hidden dimensions of the Transformer decoder and FFN in model $\Xi$, respectively.
	With the pre-trained draft and verify models, the inference process of the considered system comprises the following procedures.
	
	1) \textbf{Input Sequence Uploading}: 
	For each inference task $k$, the corresponding input sequence is first transmitted from the requesting user to the SBS via wireless uplink. 
	
	2) \textbf{Draft Phase}:
	Based on the received input sequences, the draft model at the SBS adopts AD to generate $l$ draft tokens for each task $k$.
	The inference procedure with AD consists of two stages, i.e., a  \textit{prefill stage} and a \textit{decoding stage}.
	In the \textit{prefill stage}, the draft model processes the input sequence of each task $k$ through its $J_{\mathrm{d}}$ cascaded decoder blocks to generate the first draft token.
	The input of each decoder block $j$ for task $k$ is specified by
	$\mathbf{C}^j\in\mathcal{R}^{I_k\times h_1^{\rm d}}$, 
	where  
	$\mathbf{C}^j = \{\mathbf{c}_{i}^j:1\le i\le I_k\}$ and $\mathbf{c}_{i}^j$ is the embedding of decoder block $j$ for $i$-th input token.
	The query, key, and value of $\mathbf{C}^j$ are computed by  
	\begin{align} \label{eq:f11}
		\mathbf{Q}^j=\mathbf{C}^j \mathbf{W}_Q^j; ~
		\mathbf{K}^j=\mathbf{C}^j\mathbf{W}_K^j; ~
		\mathbf{V}^j=\mathbf{C}^j\mathbf{W}_V^j.
	\end{align}

	With $\mathbf{Q}^j$, $\mathbf{K}^j$ and $\mathbf{V}^j$,
	the output of the MHA is given by
	\begin{align} \label{eq:f21}
		\mathbf{C}_{\rm{out}}^j = f_{\rm{softmax}}({{\mathbf{Q}}^j(\mathbf{K}^{j})^T}/{\sqrt{d_0}})\mathbf{V}^j\mathbf{W}_O^j+\mathbf{C}^j.
	\end{align}
	
	Using $\mathbf{C}_{\rm{out}}^j$ as the input of the FFN and residual connection, the output of decoder block $j$ is given by 
	\begin{align} \label{eq:f31}
		\mathbf{C}^{j+1} = f_{\rm{relu}}(\mathbf{C}_{\rm{out}}^j\mathbf{W}_1^j )\mathbf{W}_2^j+\mathbf{C}_{\rm{out}}^j.
	\end{align}
 
	Based on the output of the final decoder block, i.e., $\mathbf{C}^{J_{\mathrm{d}}+1}$, the draft model produces a set of conditional probability distributions $\{{q(\cdot|\mathbf{c}_{{\le i}})}: 1\le i\le I_k\}$, where $q(\cdot |\mathbf{c}_{{\le i}})$ denotes the probability distribution over the vocabulary for predicting the next token given the first $i$ input tokens, i.e., $\mathbf{c}_{{\le i}}$.
	The first draft token is sampled from ${q(\cdot|\mathbf{c}_{{\le I_k}})}$, 
	since it is conditioned on all input tokens.
	To reduce computation, the \textit{prefill stage} constructs the KV cache for each decoder block, which is reused for draft token generation during the \textit{decoding stage}.
	In the \textit{decoding stage}, the draft model utilizes and updates the KV cache to generate subsequent draft tokens.
	Let $\mathbf{t}^j\in\mathcal{R}^{1\times h_1^{\rm d}}$ denote the embedding of the newly generated token in decoder block $j$.
	The query, key, and value of $\mathbf{t}^j$ are computed as
	$\mathbf{q}^j_{\rm{new}}\!\!=\!\!\mathbf{t}^j\mathbf{W}_Q^j$,		
	$\mathbf{k}^j_{\rm{new}}\!\!=\!\!\mathbf{t}^j\mathbf{W}_K^j, \mathbf{v}^j_{\rm{new}}\!\!=\!\!\mathbf{t}^j\mathbf{W}_V^j$, respectively.
	The KV cache is then updated by
	\begin{align} \label{eq:d_kv}  
		\mathbf{K}^j\!=\! {\rm{Concat}}(\mathbf{K}^j,\mathbf{k}^j_{\rm{new}});~~
		\mathbf{V}^j\!=\!{\rm{Concat}}(\mathbf{V}^j,\mathbf{v}^j_{\rm{new}}).
	\end{align}
	
	Based on \eqref{eq:d_kv}, the output of decoder block $j$ is computed as
	\begin{align} 
		\mathbf{t}_{\rm{out}}^j &= f_{\rm{softmax}}({{\mathbf{q}}_{\rm{new}}^j(\mathbf{K}^j)^T}/{\sqrt{d_0}})\mathbf{V}^j\mathbf{W}_O^j+\mathbf{t}^j, \label{eq:d_atte}\\
		&~~~\mathbf{t}^{j+1} = f_{\rm{relu}}(\mathbf{t}_{\rm{out}}^j\mathbf{W}_1^j )\mathbf{W}_2^j+\mathbf{t}_{\rm{out}}^j. \label{eq:d_out}
	\end{align}
	
	Given that the newly generated token is the $(l\!\!-\!\!1)$-th draft token,  the draft model utilizes the output of the last decoder block, i.e., $\mathbf{t}^{J_{\rm{d}}+1}$, to generate the probability distribution for predicting $l$-th draft token $\mathbf{t}_{l}$, denoted by $q_{k,l}={q(\cdot|\mathbf{c}_{\le I_k},\mathbf{t}_{\le l-1})}$.
	Once $l$ draft tokens are generated, the SBS transmits them and the corresponding probability distributions to the MBS via the uplink fronthaul link.
	
	3) \textbf{Verification Phase}: 
	The verify model at the MBS evaluates $l$ draft tokens in parallel.
	Similar to the \textit{prefill stage} of AD,
	the verify model generates probability distributions over the vocabulary for each position of the current input, including the input sequence and $l$ draft tokens.
	The set of probability distributions is denoted by $\{p(\cdot|\mathbf{c}_{ \le i}):1\le i\le I_k\} \cup \{p(\cdot|\mathbf{c}_{\le I_k},\mathbf{t}_{\le l}:1\le l \le l\}$.
	Each draft token $\mathbf{t}_{l}$ is evaluated using the corresponding probability distribution, i.e., $p_{k,l}=p(\cdot|\mathbf{c}_{\le I_k},\mathbf{t}_{\le l-1})$.
	Specifically, each draft token $\mathbf{t}_{l}$ is accepted with probability $\min (1,p_{k,l}/q_{k,l})$.
	In addition to the accepted tokens, the generated output tokens include a bonus token.
	The bonus token is either the correction token for the first unaccepted draft token or the reward token when all the draft tokens are accepted based on the speculative sampling methods \cite{leviathan2023fast}. 
	According to \cite{leviathan2023fast,sadhukhan2024magicdec}, the average number of generated output tokens depends on the number of draft tokens, i.e., speculation length $l$, and the average token acceptance rate, denoted by $\alpha\in(0,1)$. 
	The acceptance rate is determined by the distributional similarity between the draft model and the verify model.
	A higher acceptance rate and a longer speculation length contribute to generating more output tokens.
	With $l$ and $\alpha$, the average number of generated output tokens after one verification is given by
	\cite{leviathan2023fast} 
	\begin{align} \label{eq:ol}
		L=\frac{1-\alpha^{1+l}}{1-\alpha}.
	\end{align}
	
	Then, the MBS  transmits $L$ output tokens to the SBS over the downlink fronthaul link.
	Similar to the draft model, the KV-cache mechanism is adopted for the verify model to reduce computation.
	For both the draft and verify models, the rejected draft tokens are removed from the KV cache after the verification phase to save memory resources \cite{butler2024pipeinfer}.

	4) \textbf{Output Token Downloading}:
	Upon receiving $L$ output tokens, the SBS immediately delivers them to users via the wireless downlink.
	For clarity, we refer to a draft phase followed by a verification phase as a decoding step.
	The decoding step is repeated until the complete output token sequence for task $k$ is generated.
	Note that, at the next decoding step, the draft model takes the bonus token as input and iteratively generates $l$ new draft tokens.
	The verify model then takes the bonus token together with the newly generated $l$ draft tokens as input to produce the next output tokens.

	\subsection{Pipeline-enabled Serving Mechanism with SD} \label{sec:pipe}
		\begin{figure}[t]
		\centering
		\includegraphics[width=0.45\textwidth]{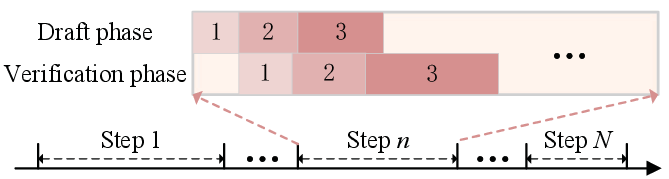}\\
		\caption{Illustration of the total latency of pipeline-enabled LLM inference.}
		\label{fig:pipeline}
		\vspace{-10pt} 
	\end{figure}
	To improve resource utilization and reduce serving latency, we propose a pipeline-enabled serving mechanism with SD, in which the draft phase and the verification phase are orchestrated in parallel across different tasks.
	Specifically, all inference tasks are first partitioned into $M$ batches,
	which are then processed by the MBS and SBS at the batch level.
	For simplicity, we denote the set of batches as $\mathcal{M}=\{1,2,\cdots,M\}$.
	Since batching requires the input of all tasks in a batch to have the same shape, the input sequence of each task in batch $m$ is right-padded to a fixed length $I_m = \mathop {\max }\limits_k \{x_k^mI_k\}$, where $x_{k}^m\in\{0,1\}$ indicates whether the inference task $k$ is assigned to batch $m$.
	For ease of presentation, let $\bm{x}=\{x_k^m:\forall m\in\mathcal{M},k\in\mathcal{K}\}$ and $N$ represent all tasks' batching scheduling decisions and total decoding step number for completing all tasks, respectively.
	
	Then, all batches are processed sequentially in a two-stage pipeline at each decoding step, with the draft phase as the first stage and the verification phase as the second stage, as illustrated in Fig.~\ref{fig:pipeline}.
	During the first decoding step, all batches participate in the pipeline. 
	In subsequent decoding steps, only unfinished batches remain in the pipeline, enabling early completion for some tasks.
	The total decoding step number $N$ depends on the latest completed batch, i.e., $N = \mathop {\max }\limits_m \{n_m\}$, where $n_m$ denotes the number of decoding steps required to complete batch $m$.
	The value of $n_m$ is related to the maximum expected output token length among the tasks in batch $m$, and the number of generated output tokens for each task per decoding step, i.e., $L$, in \eqref{eq:ol}.
	The maximum expected number of output tokens for batch $m$ is expressed as ${O_m} = \mathop {\max }\limits_k \{x_k^mO_k^m\}$.
	Given $O_m$ and $L$, the decoding step number of batch $m$ is given by 
	\begin{align} \label{eq:step_n}
		{n_m} = \left\lceil {{{{O_m}}}/{L}} \right\rceil,
	\end{align}
	where $\left\lceil {\cdot} \right\rceil$ is the ceiling function.

	\subsection{Inference Cost Model} \label{sec:cost}
	This subsection presents the inference cost in terms of memory consumption and computational cost.
	
	1) \textbf{Memory Cost}: 
	The memory consumption of LLM inference with SD stems from the execution of both the draft and verify models.
	For the draft model, memory usage is primarily attributed to model weights and KV-cache.
	The model parameters are predominantly determined by the weight matrices of its $J_{\rm{d}}$ decoder blocks, namely $\mathbf{W}_Q,\mathbf{W}_K,\mathbf{W}_V, \mathbf{W}_O,\mathbf{W}_1$, and $\mathbf{W}_2$.
	Accordingly, the total parameter count of the draft model is formulated as $J_{\rm{d}}(4 (h_1^{\rm d})^2 + 2 h_1^{\rm d} h_2^{\rm d})$, where the expression in parentheses represents the number of parameters per decoder block \cite{sheng2023flexgen}.
	Assuming each parameter is stored with FP16 precision (2 bytes),  the memory consumption for loading the draft model parameters is given by 
	\begin{align} \label{eq:memory_model}
		\Gamma_{\rm{p}}^{\rm{d}}= J_{\rm{d}}(8 (h_1^{\rm d})^2 + 4 h_1^{\rm d} h_2^{\rm d}).
	\end{align}
	
	In addition to parameter storage, the KV-cache is a major contributor to memory consumption.
	For each batch $m$ at any decoding step $n$,  the memory usage of the KV-cache depends on the total number of tokens requiring KV-cache storage.
	These tokens include the input sequence of length $I_m$, all output tokens generated prior to decoding step $n$, and the draft tokens produced at decoding step $n$ for future output prediction.
	For each task $k$ in batch $m$, the combined number of generated output tokens and current draft tokens does not exceed the maximum expected output length of the batch, i,e., $O_m$.  
	Thus, the number of tokens requiring KV-cache  for each task in batch $m$ is upper-bounded by $(I_m + O_m)$.
	With FP16 precision, the maximum memory consumption for KV-cache per task  in batch $m$ is ${4} J_{\rm{d}} h_1^{\rm d}({I_{m}} + {O_{m}})$, where ${4} J_{\rm{d}} h_1^{\rm d}$
	represents the memory usage for caching each token's key–value pair \cite{sheng2023flexgen}.
	In addition, the number of tasks in batch $m$ is expressed as $\sum_{k=1}^K x_k^m$.
	Therefore,  the peak memory consumption of KV-cache for the draft model when processing  batch $m$ is given by
	\begin{align}  \label{eq:memory_kv}
		\Gamma_{{\rm{kv}},m}^{\rm{d}}=\sum\nolimits_{k = 1}^K {4}J_{\rm{d}} {h_1^{\rm d}({I_{m}} + {O_{m}})x_{k}^m}. 
	\end{align}

	Based on \eqref{eq:memory_model} and \eqref{eq:memory_kv}, the peak memory consumption of the draft model for batch $m$ is expressed as $\Gamma^{\rm{d}}_m=\Gamma_{\rm{p}}^{\rm{d}}+\Gamma_{{\rm{kv}},m}^{\rm{d}}$.
	Note that the memory analysis does not account for the memory consumption of the verify model.
	This is because the	MBS is typically equipped with more powerful computational resource and has more memory space than the SBS \cite{8894168}.

	2) \textbf{Computational Cost}: 
	During the inference procedure, the computational cost is defined as the floating-point operations (FLOPs) consumed by the draft and verify models across all decoding steps. 
	For batch $m$, each decoding step involves  $l$ sequential forward passes through the draft model to autoregressively generate $l$ draft tokens, followed by a single forward pass through the verify model to evaluate these draft tokens.
	Notably, for the draft model, the  initial forward pass during the first decoding step corresponds to the \textit{prefill stage}, while the subsequent forward passes belong to the \textit{decoding stage}, as discussed in Section \ref{subsec:inference_model}.
	Similar to many existing works\cite{zhang2024beyond, sheng2023flexgen},
	we focus on FLOPs incurred by matrix multiplications.
	Given that matrix multiplication between $\mathbf{A} \in \mathcal{R}^{a \times b}$ and $\mathbf{B} \in \mathcal{R}^{b \times c}$ requires $2abc$ FLOPs, the FLOPs incurred by the draft model during the \textit{prefill stage} for batch $m$ are given by
	\begin{align} \label{eq:flops_1}
	F_1=J_{\rm d}(8 I_m(h_1^{\rm d})^2 +4I_m^2h_1^{\rm d}+ 4I_mh_1^{\rm d} h_2^{\rm d}),
	\end{align} 
	where $(8 I_m(h_1^{\rm d})^2 +4I_m^2h_1^{\rm d})$ represents the FLOPs required for computing $\mathbf{Q}^j,\mathbf{K}^j,\mathbf{V}^j$, and $\mathbf{C}_{\rm{out}}^j$, while $4I_mh_1^{\rm d} h_2^{\rm d}$ accounts for the FLOPs to compute $\mathbf{C}^{j+1}$ in each decoder block $j$.
	
	Similarly,  the FLOPs for the draft model to process batch $m$ during the \textit{decoding stage}  are expressed as
	\begin{align} \label{eq:flops_2}
	F_2=J_{\rm d}(8(h_1^{\rm d})^2 +4(I^{\rm{kv,d}}+1)h_1^{\rm d}+ 4h_1^{\rm d} h_2^{\rm d}),
	\end{align} 
	where $(8(h_1^{\rm d})^2 +4(I^{\rm{kv,d}}+1)h_1^{\rm d})$ represents FLOPs for computing $\mathbf{k}^j_{\rm{new}},\mathbf{q}^j_{\rm{new}},\mathbf{v}^j_{\rm{new}}$ and $\mathbf{t}_{\rm{out}}^j$,  and $4h_1^{\rm d} h_2^{\rm d}$  corresponds to the FLOPs for computing $\mathbf{t}^{j+1}$, and $I^{\rm{kv,d}}$ denotes the length of KV-cache for each decoder block.
	Based on \eqref{eq:flops_1} and \eqref{eq:flops_2}, the FLOPs required by $i$-th forward pass of the draft model at decoding step $n$ for each task in batch $m$ are given by
		\begin{align} \label{eq:flops_d}
		F_{i,n,m}^{\mathrm{d}} \!=\!
		\begin{cases}
			\!\!4J_{\mathrm{d}}h_1^{\mathrm{d}}I_m \big[
			2 h_1^{\mathrm{d}} 
			\!+\!  I_m
			\!+\!   h_2^{\mathrm{d}}
			\big], & \!\!\text{if } n\!\!=\!\!1, i\!\!=\!\!1, \\
			\!\!4J_{\mathrm{d}}h_1^{\mathrm{d}} \big[
			2h_1^{\mathrm{d}}
			\!+\!  I^{\mathrm{kv,d}}_{i,n,m} \!+\! 1 
			\!+\!  h_2^{\mathrm{d}} 
			\big], & \!\!\text{else},
		\end{cases}
	\end{align}
	where $I^{\mathrm{kv,d}}_{i,n,m}=I_m+((n\!-\!1)L\!-\!
	1)\!+\!i\!-\!1$ is the current KV-cache length of $i$-th forward pass to process batch $m$ at decoding step $n$.
	Here, the first term represents the initial input length, 
	the second term accounts for all output tokens  generated prior to  the bonus token generated in the $(n - 1)$-th step,
	and the last term corresponds to the $(i \!-\! 1)$ draft tokens generated before the $i$-th forward pass within step~$n$.	
 
	Similar to the draft model, we analyze the FLOPs of the verify model at each decoding step.
	Recalling Section~\ref{subsec:inference_model}, when processing batch $m$, the verify model at the first decoding step follows the same computational procedure as the \textit{prefill stage}, with an input of length  $(I_m + l)$.
	At the subsequent decoding steps, the verify model follows a computational procedure similar to the \textit{decoding stage}, with an input of length $(1+ l)$. 
	Thus, the FLOPs required by the verify model at any decoding step can be derived by adjusting the input length and substituting the verify model’s hyperparameters, i.e., $h_1^{\mathrm{v}},h_2^{\mathrm{v}}$ and $J_{\mathrm{v}}$, into the FLOPs expressions of the \textit{prefill} and \textit{decoding stages}, i.e., \eqref{eq:flops_1} and \eqref{eq:flops_2}.
	The FLOPs required by the verify model at decoding step $n$ for each task in batch $m$ are given by	
	\begin{align}  \label{eq:flops_v}
		F_{n,m}^{\mathrm{v}} \!\!\!=\!\!\!
		\begin{cases}
			\!\!4 J_{\mathrm{v}} h_1^{\mathrm{v}} (I_m \!+\! l) 
			\big[
			2 h_1^{\mathrm{v}} \!+\! I_m \!\!+\!\! l + h_2^{\mathrm{v}}
			\big], & \text{if } n \!\!=\!\! 1, \\[2pt]
			\!\!4 J_{\mathrm{v}} h_1^{\mathrm{v}}\! (1 \!+\! l) 
			\big[
			2 h_1^{\mathrm{v}} \!\!+\!\! I^{\mathrm{kv},\mathrm{v}}_{n,m} \!\!+\!\! l \!\!+\! 1 \!\!+\!\! 	h_2^{\mathrm{v}}
			\big], & \text{else},
		\end{cases}
	\end{align}
	where $I^{\mathrm{kv,v}}_{n,m}\!=\! I_m \!+\! (n - 1)L \!-\! 1$ is the current KV-cache length of batch $m$ at decoding step $n$.
	The current KV-cache includes the input sequence of length $I_m$ and all output tokens preceding the bonus token generated in step $(n - 1)$, with a total length of $\bigl((n-1)L - 1\bigr)
	$.	

	\subsection{Serving Latency Model}
	In practice, improving quality of service (QoS) under memory and computational cost constraints is essential in LLM serving systems. 
	This work measures QoS using total serving latency,  encompassing communication and inference latencies.

	1) \textbf{Communication Latency}:
	The wireless communication process involves the uploading of input sequences from users to the SBS via wireless uplink, and the downloading of output tokens from the SBS to users through wireless downlink. 
	Notably, the downlink transmission latency is  negligible, as the SBS operates with higher transmit power than users, enabling a high downlink transmission rate. 
	Moreover, output tokens are streamed to users immediately after each verification phase, without waiting for the complete output sequences.
	This streaming transmission mechanism, widely  adopted in LLM serving systems, effectively reduces downlink transmission load \cite{li2024eloquent}.
	Hence, this work focuses on the uplink transmission latency.
	Similar to \cite{chen2023knowledge,zhu2025joint}, the orthogonal frequency division multiple access protocol is employed to upload users' input sequences with total uplink bandwidth $B_{\rm{w}}$ Hz.
	Let $\bm{w}=\{w_{k}:\forall k\in\mathcal{K}\}$ 
	denote the system-wide uplink bandwidth resource allocation, where $w_k$ represents the bandwidth ratio allocated to task $k$.
	For each task $k$, the uplink channel gain and transmit power are represented by $g_{k}$ and $p_{k}$, respectively.
	The achievable uplink rate of task $k$ is expressed as $r_{k} = w_{k}B_{\rm{w}}{\log _2}\left( {1 + {{p_{k}g_{k}}}/{{{\sigma ^2}}}} \right)$, where  $\sigma ^2$ is the noise power.
	Thus, the uploading latency of the input sequence for task $k$ is given by
	\begin{align} \label{eq:ul_latency_k}
		T_{k,{\rm{com}}}=\frac{{\lambda {I_{k}}}}{{r_{k}}}
		=\frac{\lambda I_k}{w_kB_{\mathrm{w}}\log _2\left( 1+{p_kg_k}/{\sigma ^2} \right)},
	\end{align}
	where ${\lambda {I_k}}$ is the data size (in bits) of the input sequence for task $k$, and $\lambda$ denotes the number of bits required to store embedding data per input token.
	For instance, under FP16 precision, $\lambda=16(h_1^{\rm d}+h_1^{\rm v})$, where $16h_1^{\rm d}$ and $16h_1^{\rm v}$ denotes the embedding size per token for the draft model and the verify model, respectively.

	Since the batch processing requires
	input sequences of all tasks to be transmitted to the SBS before inference, the communication latency determined by the latest-arriving task is given by
	\begin{align} \label{eq:ul_latency}
		T_{{\rm{com}}} = \mathop {\max }\limits_{k \in \mathcal{K}} \left\{T_{k,{\rm{com}}}\right\}.
	\end{align}

	2) \textbf{Inference Latency}:
	Once the input sequences for all tasks are uploaded, the SBS and MBS collaboratively execute the inference process illustrated in Fig.~\ref{fig:pipeline}.
	The inference latency is defined as the cumulative processing latency of both the SBS and MBS across all decoding steps. 
	The processing latency at each decoding step corresponds to the execution latency of a two-stage pipeline.
	The latency of the first stage encompasses the runtime of the draft model for generating draft tokens and the uplink fronthaul transmission delay for sending draft tokens to the MBS.
	For the second stage, the corresponding latency
	includes the verify model's runtime and the downlink fronthaul transmission latency for delivering verification results to the SBS.
	Note that the latency of uplink and downlink fronthaul transmission latency is negligible.
	This is because both the uploaded draft tokens and the downloaded verification results are small in size and transmitted over high-speed fronthaul links. 
	For example, when running LLaMA-2.7B on a Jetson Orin Nano, the average per-token inference latency is approximately 30\,ms, while transmitting the data of a draft token, including the token id and its probability distribution, over a fronthaul link with a rate of 10\,Gbps takes only about 10\,$\mu$s in FP16 precision~\cite{jetson_tutorial}.
	Hence, we use the runtime of the draft and verify models to characterize the processing latency at each decoding step.
	As found in many existing works \cite{shi2022multiuser,cang2024joint,xu2023smdp}, there exists an approximately linear relationship between the model runtime and the batch size for different models on a given hardware platform.
	However, to the best of our knowledge, no prior work characterizes the relationship among model runtime, batch size, and FLOPs. 
	To bridge this gap, we model this relationship as follows
		\begin{align} \label{eq:latency_b2}
		T(b,F) = c_1 F b + c_2,
	\end{align}
	where $b$ represents the batch size, $F$ denotes the required FLOPs for a forward pass of models, and $c_1$,  $c_2$ are fitting coefficients that depend on the specific hardware.
		\begin{figure}[t]
		\centering
		\includegraphics[width=0.45\textwidth]{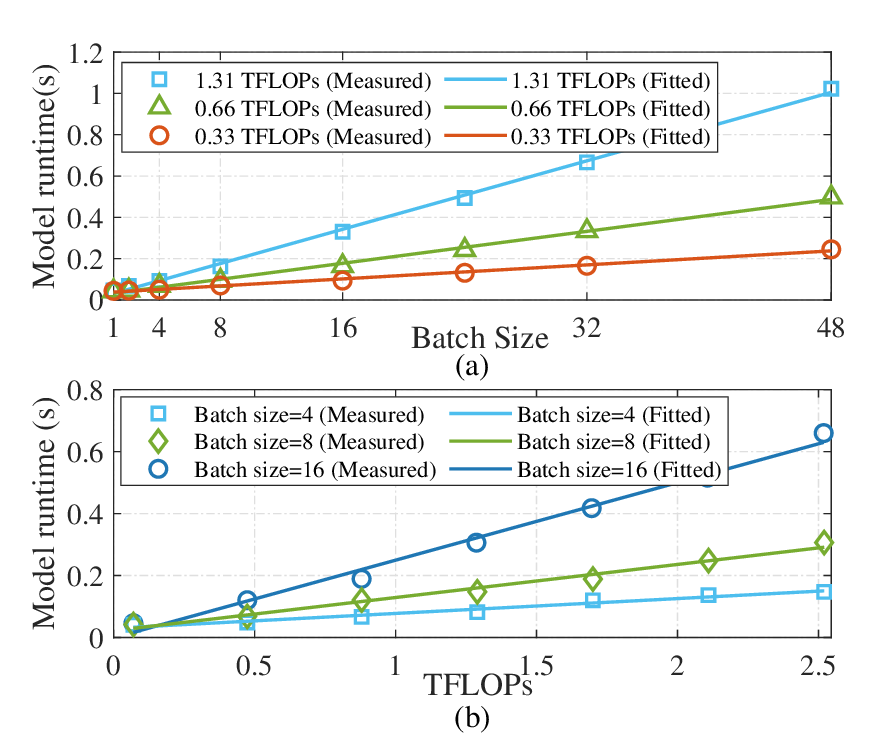}\\
		\setlength{\abovecaptionskip}{1.0pt}
		\caption{Model runtime of LLaMA-7B versus batch size and FLOPs on NVIDIA RTX4500.}
		\label{fig:fit_l}
		\vspace{-10pt} 
	\end{figure}
	
	To validate the relationship, we conducted experiments with LLaMA-7B on an NVIDIA RTX 4500, with the model details provided in the experimental setup in Section \ref{sec:simulation}.
	Fig. \ref{fig:fit_l} presents the measured runtime as a function of (a) batch size and (b) FLOPs, along with linear regression fits.
	The results confirm that model runtime scales near-linearly with both batch size and FLOPs.
	Therefore, we adopt \eqref{eq:latency_b2}  to characterize the runtime of each forward pass in LLM inference.
	Accordingly, the runtime of the verify model to process batch $m$ at any decoding step $n$ is given by 
	\begin{align} \label{eq:v_latency}
		{T_{n,m}^{\rm{v}}}= c_1^{\rm{v}} F_{n,m}^{\rm{v}}  \sum\nolimits_{k = 1}^K x_k^m + c_2^{\rm{v}},
	\end{align}
	where $ \sum_{k = 1}^K x_k^m$ 
	is the size of batch $m$, and $c_1^{\rm v}, c_2^{\rm v}$ are fitting coefficients specific to the GPU at the MBS.
	
	Similarly, the runtime of the draft model for processing batch $m$ at decoding step $n$ is given by 
	\begin{align} \label{eq:d_latency}
		T_{n,m}^{\rm d} = \sum\nolimits_{i=1}^{l} (c_1^{\rm{d}} F_{i,n,m}^{\rm{d}} \sum\nolimits_{k = 1}^K x_k^m + c_2^{\rm{d}}),
	\end{align}
	where $(c_1^{\rm{d}} F_{i,n,m}^{\rm{d}} \sum\nolimits_{k = 1}^K x_k^m + c_2^{\rm{d}})$ represents the runtime of the draft model to generate the $i$-th draft token for each task in batch $m$ at decoding step $n$.
	$c_1^{\rm{d}}$ and $c_2^{\rm{d}}$ are fitting coefficients related to the GPU deployed at the SBS.
	Note that the required FLOPs for the draft and verify models to process batch $m$ at each decoding step $n$, i.e., $F_{i,n,m}^{\rm{d}}$ and $F_{n,m}^{\rm{v}}$, can be found in Section~\ref{sec:cost}.	
	 
	For the two-stage pipeline at decoding step $n$, the latencies of the first and second stages to process batch $m$ are  $T_{n,m}^{\rm d}$ and $T_{n,m}^{\rm v}$, respectively. 
	According to the classical pipeline scheduling theory \cite{johnson1954optimal}, 
	the second stage of batch $m$ can only start after 
	both the first stage of batch $m$ 
	and the second stage of batch $(m-1)$ have completed.
	Let $C^{\rm{d}}_{n,m}=\sum_{i=1}^m T^{\rm{d}}_{n,i}$ denote the completion time for processing the first $m$ batches in the first stage at decoding step $n$.
	Correspondingly, the completion time of the first $(m-1)$ batches through both stages of decoding step $n$ is defined as $C_{n,m-1}$.	
	Note that the initial condition is $C_{n,0}=0$ for each decoding step $n$.
	Thus, the start time of each batch $m$ at the second stage of decoding step $n$ is expressed as $\max\left\{C^{\rm{d}}_{n,m}, C_{n,m-1}\right\}$.
	Given the processing latency of batch $m$ at the second stage, i.e., $T_{n,m}^{\rm v}$, the latency of processing the first $m$  batches at decoding step $n$ is given by 
	\begin{equation} \label{eq:time}
		C_{n,m} =  \max\left\{C^{\rm{d}}_{n,m}, C_{n,m-1}\right\} + T^{\rm{v}}_{n,m}.
	\end{equation}

	The total execution latency for decoding step $n$ equals the latency of processing all active batches of decoding step $n$, which is given by
	\begin{equation} \label{eq:latency_infer_batch}
		T_n = C_{n,M_n},
	\end{equation}
	where $M_n = |\mathcal{M}_n|$ is the number of active batches at decoding step $n$, 
	with $\mathcal{M}_n = \{m : n_m \geq n, \forall m \in \mathcal{M}\}$ 
	denoting the set of active batches whose required decoding steps, i.e., $n_m$  are not less than $n$.
		
	Then, we obtain the inference latency by summing the pipeline execution latencies across all decoding steps, i.e.,
	\begin{equation} \label{eq:latency_inf}
		T_{\rm{inf}} = \sum\nolimits_{n=1}^{N} T_n.
	\end{equation} 
	
	Thus, the total serving latency for all tasks is given by
	\begin{align}
		{T} = T_{\rm{inf}} + T_{\rm{com}}.
	\end{align}

	\subsection{Problem Formulation} \label{subsec:problem_formulation}

		This work aims to minimize the total serving latency of all tasks, i.e., $T$.
		Specifically, we jointly optimize the decisions of batching scheduling, the number of batches, speculation length, and communication resource allocation under the limited memory and communication resources at the edge.
		The optimization problem is formulated as follows
		\begin{align}
			\mathcal{P}:~~&\min_{\left\{ 
				\bm{x},M,l,\bm{w} \right\}} T  \label{prob:P}\\
			\text{s.~t.~~} 
			&\sum\nolimits_{k = 1}^K {w_{k } \le 1},  \label{cons:P_1}\tag{\theequation a}\\
			&{\Gamma_m^{{\rm{d}}}} \le {\Gamma_{\rm{s}}}, \forall m,\label{cons:P_2}\tag{\theequation b} \\
			&\sum\nolimits_{m = 1}^M {x_{k }^m = 1}, \forall k, \label{cons:P_3}\tag{\theequation c}\\
			&x_{k}^m \in \{0,1\} ,\forall k,m, \label{cons:P_4}\tag{\theequation d} \\
			&0 \le w_{k} \le 1,\forall k, \label{cons:P_5}\tag{\theequation e}\\
			&1 \le {l} \le {l_{\rm{max}}},\forall {l} \in {\mathbb{Z}^+},\label{cons:P_6}\tag{\theequation f}\\
			&M\le K,\forall M \in {\mathbb{Z}^+},\label{cons:P_7}\tag{\theequation g}
		\end{align}
		where 
		\eqref{cons:P_1} guarantees that the total wireless uplink bandwidth allocated  to tasks is not greater than the available bandwidth resources.
		Furthermore, \eqref{cons:P_2} ensures that the memory consumption of the draft model
		for each batch $m$ cannot exceed the available memory capacity of the SBS, i.e., $\Gamma_{\rm s}$.
		\eqref{cons:P_3} is to ensure that each task must be assigned to a batch for processing.
		Meanwhile, \eqref{cons:P_4} and \eqref{cons:P_5} define the feasible value ranges for batching and bandwidth allocation decisions, respectively.
		\eqref{cons:P_6} ensures that the speculation length does not exceed the maximum threshold $l_{\rm{max}}$, thereby preventing excessive computational resource waste due to verification failures.
		\eqref{cons:P_7}  indicates that the number of batches cannot surpass the task number.
		Problem $\mathcal{P}$ is a mixed-integer nonlinear programming problem, which is generally NP-hard \cite{du2017computation}.
		The difficulty in solving problem $\mathcal{P}$ mainly arises from the strong coupling among decision variables and 
		the complexity of the objective function, which lacks an explicit closed form due to the recursive nature of inference latency.
		To address these challenges, we propose an efficient algorithm for communication resource allocation, batching, and speculation length control to solve problem  $\mathcal{P}$.

	
	\section{Efficient Resource Allocation, Batching and Speculation Length Control Policy}  \label{sec:algor}
	In this section, we first decompose the communication resource allocation subproblem from problem $\mathcal{P}$, and derive its closed-form optimal solution.
	Then, we  develop a dynamic programming-based algorithm to obtain efficient batching and speculation length control strategies.
	
%
%

	\vspace{-0.6em}
	\subsection{Optimal Communication Resource Allocation}\label{sec:policy}
	In problem $\mathcal{P}$, the communication resource allocation variable, i.e., $\bm{w}$, only impacts 
	the communication latency and the start time of inference procedure, and does not affect the inference latency.
	Therefore, we decouple the communication resource allocation subproblem from problem $\mathcal{P}$, i.e., 
	\begin{align}
		\mathcal{P}_1:~~&\min_{ 
			\bm{w} } T_{{\rm{com}}}= \mathop {\max }\limits_{k \in \mathcal{K}} \left\{T_{k,{\rm{com}}}\right\}
			 \label{prob:P_1}\\
		\text{s.~t.~~} 
		&\eqref{cons:P_1},\eqref{cons:P_5}. \notag
	\end{align}
		
	According to \eqref{eq:ul_latency_k}, it is obtained that the uploading latency of each task, i.e., $T_{k,{\rm{com}}}=\frac{\lambda I_k}{w_kB_{\mathrm{w}}\log _2\left( 1+{p_kg_k}/{\sigma ^2} \right)}$, monotonically decreases with the increase of the allocated bandwidth ratio $w_k$. 
	If a task completes the upload of its input sequence earlier than others, a portion of its allocated bandwidth can be reallocated to slower tasks, thereby reducing the overall communication latency determined by the slowest task.
	The reallocation process continues until all tasks complete uploading their input sequences simultaneously.
	Therefore, the optimal solution to $\mathcal{P}_1$ is achieved when  bandwidth allocation ensures that all tasks finish uploading at the same time.
	Hence, the optimal communication resource allocation policy satisfies
	\begin{equation} \label{eq:bandw1}
		\left\{ {\begin{array}{*{20}{c}}
				{{w_k} = \frac{{\lambda {I_k}}}{{t_{\rm{com}}^*{B_{\rm{w}}}{{\log }_2}\left( {1 + {{{p_k}{g_k}}}/{{{\sigma ^2}}}} \right)}},\forall k,}\\
				{\sum\nolimits_{k = 1}^K {{w_k}}  = 1},
		\end{array}} \right.
	\end{equation}
	where $t_{\rm{com}}^*$ is the optimal communication latency.

	
	By solving the linear equation in \eqref{eq:bandw1},  we obtain the optimal communication latency, i.e., 
	$t_{\rm{com}}^*=\sum\nolimits_{k = 1}^K {\frac{{\lambda {I_k}}}{{{B_{\rm{w}}}{{\log }_2}\left( {1 + {{{p_k}/{g_k}}}{{{\sigma ^2}}}} \right)}}} $.
	By substituting $t_{\rm{com}}^*$ into \eqref{eq:bandw1}, the optimal communication resource allocation of each task $k$ is given by
	\begin{align} \label{eq:opt_w}
	 w_k^* = \frac{{\frac{{ {I_k}}}{{{{\log }_2}\left( {1 + {{{p_k}{g_k}}}/{{{\sigma ^2}}}} \right)}}}}{{\sum\nolimits_{k' = 1}^K {\frac{{{I_{k'}}}}{{{{\log }_2}\left( {1 + {{{p_{k'}}{g_{k'}}}}/{{{\sigma ^2}}}} \right)}}} }}.
	\end{align}	 
	\begin{rem}
	From \eqref{eq:opt_w}, the allocated communication resource of each task $k$ is monotonically decreasing with its transmit power $p_k$ and uplink channel gain $g_k$.
	It means the tasks with low transmit power and poor channel condition should be allocated by more communication resources.
	\end{rem}	
	
	\subsection{Optimal Batching and Speculation Length}\label{sec:policy batching}
	According to the formulation of problem $\mathcal{P}$,  batching and speculation length decision variables do not coupled with the communication resource allocation variables, i.e., $w_k$, and only impact inference latency, i.e., $T_{\rm{inf}}$.
	Therefore, the batching and speculation optimization subproblem can be decoupled from problem $\mathcal{P}$. 
	However, it is difficult to directly solve the subproblem 
	due to the inherent coupling between these decision variables in the model runtime, i.e., $T_{n,m}^{\rm{v}}$ and $T_{n,m}^{\rm{d}}$, which collectively determine the inference latency.
	To efficiently address the subproblem, we first optimize the batching decisions, i.e., batch scheduling variables $\{x_k^m\}$ and the number of batches $M$, under any given speculation length $l$.
	Thus, the batching optimization subproblem is expressed as
	\begin{align}
		\mathcal{P}_2:&\min_{\left\{ 
			\bm{x},M \right\}} \!T_{\rm{inf}}
			 = \sum\nolimits_{n=1}^{N} C_{n,M_n}
			  \label{prob:P_2}\\
		\text{s.~t.~~} 
		&\eqref{cons:P_2},\eqref{cons:P_3},\eqref{cons:P_4},\eqref{cons:P_7}. \notag
	\end{align}

	Solving problem~\(\mathcal{P}_2\) is challenging for three reasons: 
	1) the variable batch number $M$ introduces dynamic dimensionality in the scheduling variables \(\{x_k^m\}\), 
	2) the recursive nature of the formulation for	$C_{n,M_n}$ complicates direct optimization,
	and 3) the varying number of active batches $M_n$ across decoding steps further complicates the objective function.
	To address these difficulties, we first adopt a conservative approach by setting a maximum possible output length for each task, i.e., $O_k=O_{\rm max}, \forall k \in \mathcal{K}$.
	Under this simplification, the active batch set at each decoding step $n$ becomes identical to the complete batch set, i.e., $\mathcal{M}_n=\mathcal{M}$,  since assigning  $O_{\rm max}$ to all tasks ensures that each batch undergoes the same number of decoding steps, as defined in \eqref{eq:step_n}.
	It is worth noting that this approach is reasonable for system optimization since the output length $O_k$ of each task $k$ is uncertain and difficult to predict accurately in practice\cite{oh2024exegpt}.
	In addition, the approach ensures that the memory constraint, i.e., \eqref{cons:P_2}, is satisfied for any actual output length.


    With this simplification, we then propose a dynamic programming algorithm that efficiently solves problem $\mathcal{P}_2$. 
    Specifically, all tasks are first sorted in ascending order according to their input sequence lengths.
    For ease of presentation, the set of sorted task is denoted by $task\_list$.
    This preprocessing step is to ensure that tasks with similar input lengths are batched together, thereby reducing padding-induced computational latency. 
    Subsequently, we define a three-dimensional matrix $\varUpsilon$ with dimensions $K \times (N+1) \times 2$, where $\varUpsilon[i,0,0]$ represents the minimum  inference latency achieved by the algorithm for processing the first $i$  tasks in $task\_list$.
    To compute $\varUpsilon[i,0,0]$, we examine all possible batch boundary positions $j \in \{1,2,\cdots,i\}$, where tasks $j$ through $i$ form the final batch. 
    The size of this batch is $(i-j+1)$, and its maximum input length is $I_i$.
    Substituting $(i-j+1)$ and $I_i$ into \eqref{eq:v_latency} and \eqref{eq:d_latency}, we obtain the latencies of the verification and draft phases, respectively, for processing the batch at decoding step $n$, denoted by $T^{\rm v}_{n,\, j \le k \le i}$ and $T^{\rm d}_{n,\, j \le k \le i}$.
    According to the definition of the inference latency, i.e., \eqref{eq:latency_inf} and \eqref{eq:time}, the inference latency for processing the first $i$ tasks with each batch boundary $j$ is expressed as
    \begin{align}\label{eq:t_ij1}
    	T_{i,j} \!&=\!\! \sum\nolimits_{n=1}^N [\max\{\varUpsilon[j-1,n,0], \varUpsilon[j-1,n,1] \\
    	&+T^{\rm{d}}_{n, j\le k\le i }\}+T^{\rm{v}}_{n, j\le k\le i}],\notag
    \end{align}
    where $\varUpsilon[j-1,n,0]$ and $\varUpsilon[j-1,n,1]$ represent the completion time in the first  and second stage at decoding step $n$ for processing first $(j-1)$ tasks, as previously  stored in $\varUpsilon$. 
 
    Thus, the minimum  inference latency for processing the first $i$ tasks is given by
    \begin{align} \label{eq:rg}
    	\varUpsilon[i,0,0] = \min_{1 \leq j \leq i} T_{i,j}.
    \end{align}
    
    Let $j^* = \arg\min_{1 \leq j \leq i} T_{i,j}$ denote the batch boundary that yields this minimum latency.
    To facilitate subsequent computations, we store the completion time of the first and second stage  at decoding step $n$  for processing the first $i$ tasks, i.e.,
    \begin{align} 
    	\varUpsilon[i,n,0]&=\!\varUpsilon[j^*\!-\!1,n,0]
    	+T^{\rm{d}}_{n, j^*\le k\le i }, \label{eq:tt1}
        \end{align}
    \begin{align} 
    	\varUpsilon[i,n,1]&=\!\max\{\varUpsilon[i,n,0],\varUpsilon[j^*\!-\!1,n,1]\} \!+\!\!T^{\rm{v}}_{n, j^*\le k\le i}. \label{eq:tt2}
    \end{align}

    The core idea of the algorithm is to systematically evaluate all possible batch boundary positions for each task $i$ and select the one that minimizes the latency for processing the first $i$ tasks.
    During this process, batch boundaries that would cause memory consumption to exceed the maximum available memory of the SBS are discarded to ensure feasibility.
    By iteratively updating \eqref{eq:rg}, \eqref{eq:tt1}, and \eqref{eq:tt2} for $i \in \{1, 2, \cdots, K\} $, we obtain the minimum inference latency for processing all tasks, yielding $\varUpsilon[K,0,0]$.
    To reconstruct the corresponding batching decisions, we define a vector $S$ with dimension $K$, where $S[i]$ stores the selected batch boundary $j^*$ for task $i$. By backtracking from $S[K]$, the batching decisions, i.e., $\{x_k^m\}$ and  $M$, can be  recovered.
    For clarity, the detailed steps of the batching algorithm are summarized in \textbf{Algorithm} \ref{alg:dp} with a time complexity of $\mathcal{O}(K^2N)$. 
    This complexity arises from the $K^2$ iterations (lines 5–6) required to explore all possible batching boundaries for each task $i$. 
    Each iteration is dominated by the computation of the total inference latency for processing the first $i$ tasks (line 15), contributing an $\mathcal{O}(N)$ factor per iteration.
    Compared with exhaustive search over all batching decisions, whose complexity is $\mathcal{O}(K^K)$, the proposed algorithm significantly reduces computational overhead. 

	\begin{algorithm}[t]\small
	\caption{Dynamic Programming Algorithm for Obtaining the Batching Decisions}
	\label{alg:dp}
	\begin{spacing}{0.90}
		\begin{algorithmic}[1]
			\State \textbf{Input:} $\{I_k:k\in \mathcal{K}\}$, $l$, $O_{\rm{max}}$
			\State \textbf{Output:} The batching decision $\{x_k^m:\forall k,m\}$;
			\State Sort tasks in increasing order according to the input sequence length $I_k$, denoted by $task\_list$.
			\State Initialize $\varUpsilon=[0]_{K\times(N+1)\times2}$,  $S=[0]_{K}$, and $i=1$;
			\For{$i=1,2, \cdots, K$}
			\For {$j=1$ to $i$}
			\State Group task $j$ to $i$ in a batch.
			\State Compute the batch size, i.e., $(i-j+1)$.
			\State Obtain the maximum input length of the batch, i.e., $I_i$.
			\State Compute the memory consumption for the batch, i.e.,  $\Gamma^{\rm{d}}_{j\le k\le i}$, based on \eqref{eq:memory_model} and \eqref{eq:memory_kv}.
			\If{$\Gamma^{\rm{d}}_{j\le k\le i}>\Gamma_{\rm s}$}
			\State \textbf{continue} 
			\Comment{skip infeasible batch (out of memory)}
			\EndIf
			\State Compute verification phase latency $T^{\rm{v}}_{n, j\le k\le i}$ using \eqref{eq:v_latency}.
			\State Compute draft phase latency  $T^{\rm{d}}_{n, j\le k\le i}$ using \eqref{eq:d_latency}.
			\State Compute the total latency $temp=\sum_{n=1}^N [\max\{\varUpsilon[j-1,n,0], \varUpsilon[j-1,n,1]+T^{\rm{d}}_{n, j\le k\le i }\}+T^{\rm{v}}_{n, j\le k\le i}]$
			\If{$j==1$} 
			\State $\varUpsilon[i,0,0]=temp$.
			\State $j^*\gets1$
			\Else
			\If{$\varUpsilon[i,0,0]>=temp$} 
			\State $\varUpsilon[i,0,0]=temp$
			\State $j^*\gets j$, $S[i]=j$
			\EndIf
			\EndIf
			\State Update $\varUpsilon[i,n,0]$ and $\varUpsilon[i,n,1]$ substituting $j^*$ into \eqref{eq:tt1} and \eqref{eq:tt2}.
			\EndFor
			\EndFor
			\For{$i=K,K-1,\cdots,1$}
			\State $end\_idx\gets i$; $start\_idx\gets S[i]$
			\State pack $task\_list[start\_idx:end\_idx]$ into a batch.
			\EndFor					
			\State \Return Optimal batching decisions, i.e., $\{x_k^m\}$ and  $M$.
		\end{algorithmic}
	\end{spacing}
\end{algorithm}

	Based on \textbf{Algorithm} \ref{alg:dp}  and \eqref{eq:opt_w}, we can obtain the communication resource allocation and batching policies, denoted by $\bm{w}^*$, $\bm{x}^*$, and $M^*$, for any given speculation length $l$. 
	Thus, the total serving latency can be computed under any speculation length. 
	With $\bm{w}^*$, $\bm{x}^*$ and $M^*$, problem $\mathcal{P}$ is transformed into the following speculation length optimization problem
	\begin{align}
		\mathcal{P}_3:~&\min_{l } T_{\rm{inf}}
		\label{prob:P_9}\\
		\text{s.~t.~~} 
		&1 \le {l} \le {l_{\rm{max}}}, {l} \in {\mathbb{Z}^+}.\tag{\theequation a}
	\end{align}
	
	Directly solving problem $\mathcal{P}_3$ is challenging due to the presence of the non-covex term $l(1-\alpha^{1+l})/{(1-\alpha)}$  in $T_{\rm{inf}}$ and the ceiling operation in  $\left. N=\left. \lceil O_{\max}\left( 1-\alpha \right) /(1-\alpha ^{1+l}) \right. \right. \rceil $, which renders the objective function non-convex and non-smooth. 
	Since the feasible domain is small and contains only $l_{\rm{max}}$  positive integers, the optimal speculation length can be  obtained by enumerating all integer values of $l$.
	Note that this approach does not incur significant computational overhead, as the maximum speculation length  $l_{\rm{max}}$ typically ranges from 10 to 20 tokens in practice, which helps avoid excessive resource waste from verification failures \cite{miao2024specinfer}.

%
%
%
%
	
	\section{Simulation Results}\label{sec:simulation}
	This section evaluates the performance of the LLM serving system with SD and the proposed algorithm.
	We consider $K=100$ inference tasks generated by users who are randomly distributed within a cell of radius 400 m \cite{chen2023knowledge}.
	An SBS is deployed at the center of the cell and connected to an MBS via high-speed fiber links.
	Following \cite{chen2023exploring}, the wireless channel gain from users to the SBS is modeled as $g_{k} = g_0\varrho _k(d_k)^{-2}$, where $g_0 = -30$ dBm is the path loss constant, 
	$\varrho_k\sim\text{Exp}(1)$ represents the Rayleigh fading component, 
	and $d_k$ is the distance between the SBS and the user associated with task $k$.
	The edge servers located at the SBS and MBS are equipped with an NVIDIA GeForce RTX 3080 GPU and an NVIDIA RTX 4500 GPU, respectively, for processing inference tasks.
	For each task, the input length and expected output length are randomly chosen from the intervals $[1,I_{\rm{max}}]$ and  $[1,O_{\rm{max}}]$, respectively. 
	To process these tasks and evaluate performance, we consider three different draft-verify model pair deployment schemes across the SBS and the MBS.
	Specifically, the model pairs are (LLaMA-68M, LLaMA-7B), (LLaMA-1.1B, LLaMA-7B), and (LLaMA-1.1B, LLaMA-13B), which are commonly used for evaluating SD performance \cite{chen2025spin,miao2024specinfer}.
	The specifications of these models are summarized in Table \ref{tab:llm_settings}, while other system parameters (unless otherwise stated) are listed in Table \ref{tab:simu_settings}, following existing works \cite{miao2024specinfer,chen2023exploring,zhang2024beyond}.

	\begin{table}[t]
		\centering
		\caption{LLM settings in the simulation}
		\label{tab:llm_settings}
		\resizebox{\columnwidth}{!}{%
			\begin{tabular}{c|c|c|c}
				\hline
				\textbf{Model} & \textbf{Layer number} & \textbf{Hidden dimension} & \textbf{FFN dimension} \\
				\hline
				LLaMA-68M & 2 & 768 & 3072 \\
				LLaMA-1.1B & 22 & 2048 & 5632 \\
				LLaMA-7B & 32 & 4096 & 11008 \\
				LLaMA-13B & 40 & 5120 & 13824 \\
				\hline
			\end{tabular}
		}
	\end{table}
	\begin{table}[t]\small
	\caption{System Parameters}
	\label{tab:simu_settings}
	\begin{tabular}{p{1.08cm}|p{0.53cm}|p{1.0cm}|p{1.28cm}|p{1.0cm}|p{1.45cm}}
		\hline
		Parameter & Value & Parameter & Value& Parameter & Value\\
		\hline
		$K$ & 100 & $\sigma^2$ & -106 dBm & $c_1^{\rm{v}}$ & $\!\!\textrm{2.08}\times\!\!\textrm{10}^{-\textrm{14}}$ \\
		$O_{\rm{max}}$ & 2048 & $p_k$ & 0.2 W & $c_2^{\rm{v}}$ & $\!\!\textrm{1.28}\!\times\!\textrm{10}^{-\textrm{2}}$ \\		
		$l_{\rm{max}}$ & 10 & $\Gamma_{\rm{s}}$ & 16 GB & $c_1^{\rm{d}}$ & $\!\!\textrm{4.11}\!\times\! \textrm{10}^{-\textrm{13}}$ \\
		$I_{\rm{max}}$ & 512 & $B_{\mathrm{w}}$ & 20 MHz & $c_2^{\rm{d}}$ & $\!\!\textrm{0.56}\!\times\!\textrm{10}^{-\textrm{3}}$ \\
		\hline
	\end{tabular}

\end{table}

	\subsection{Performance of the Proposed Serving Mechanism}
	To verify the effectiveness of the proposed LLM serving system with SD,  we introduce four benchmarks:
	1) SD without pipeline (SD w/o Pipeline) \cite{zhao2024edge}: The SBS and MBS sequentially execute the draft and verification phases for each batch. 
	The batching policies remain the same as those in the proposed serving mechanism.
	2) No batching \cite{cang2024joint}: The SBS and MBS execute the proposed LLM serving mechanism without batching, i.e., processing one task at a time.
	3) Fixed speculation length (FSL) \cite{miao2024specinfer}:  The SBS and MBS serve all tasks using the proposed serving mechanism with a fixed speculation length of $l=7$, regardless of task characteristics or the number of arriving tasks. This scheme also adopts the proposed batching strategy in \ref{sec:policy batching}.
	4) AD-based serving mechanism (ADS): The verify model at the MBS processes all tasks using AD. This baseline ensures fair latency comparison by achieving the same output accuracy as the proposed mechanism, since both rely on the verify model for final output generation. 
	The batching strategy follows a heuristic approach \cite{chen2025spin} that begins with two batches, progressively increases the batch size while computing latency, and selects the optimal batch configuration when latency performance starts to degrade.
	For a fair comparison, these benchmarks employ the proposed communication resource allocation policy in Section \ref{sec:policy}.
	
	\begin{figure}[t]
		\centering
		\includegraphics[width=0.4\textwidth]{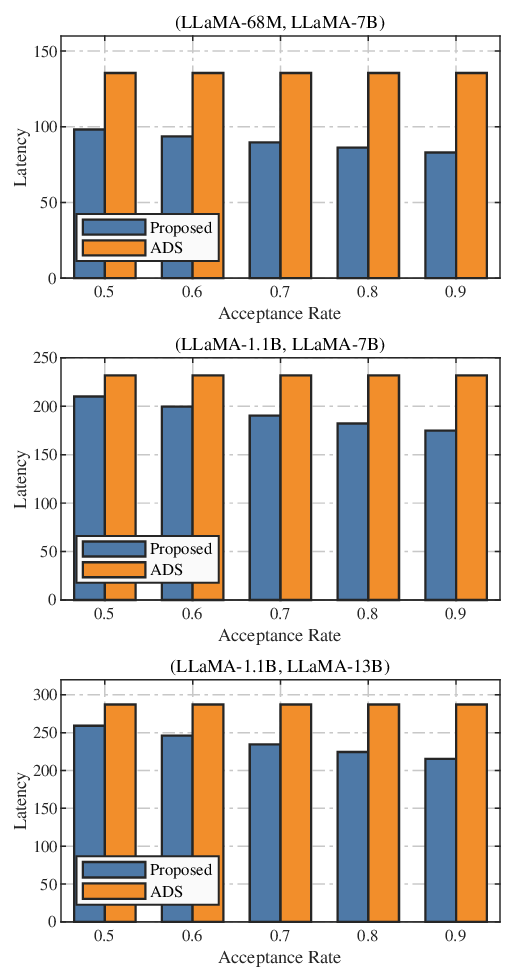}\\
		\caption{Comparison of latency of the proposed scheme and AD under different acceptance rate.}
		\label{fig:accep}
		\vspace{-12pt} 
	\end{figure}
			\begin{figure*}[t]
		\centering
		\begin{tabular}{ccc}
			\includegraphics[width=0.32\textwidth]{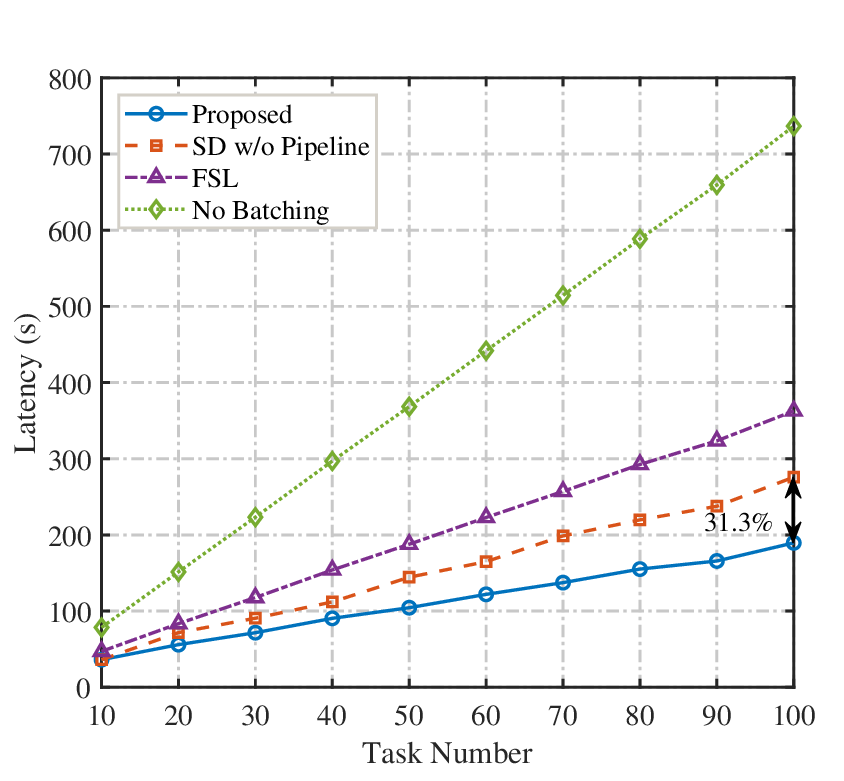} &
			\includegraphics[width=0.32\textwidth]{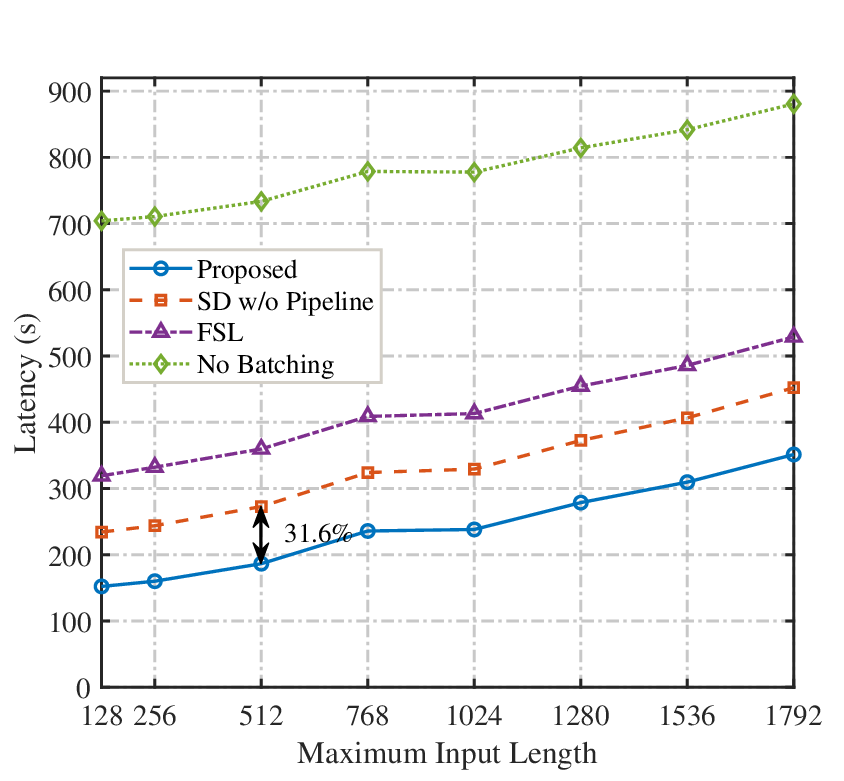} &
			\includegraphics[width=0.32\textwidth]{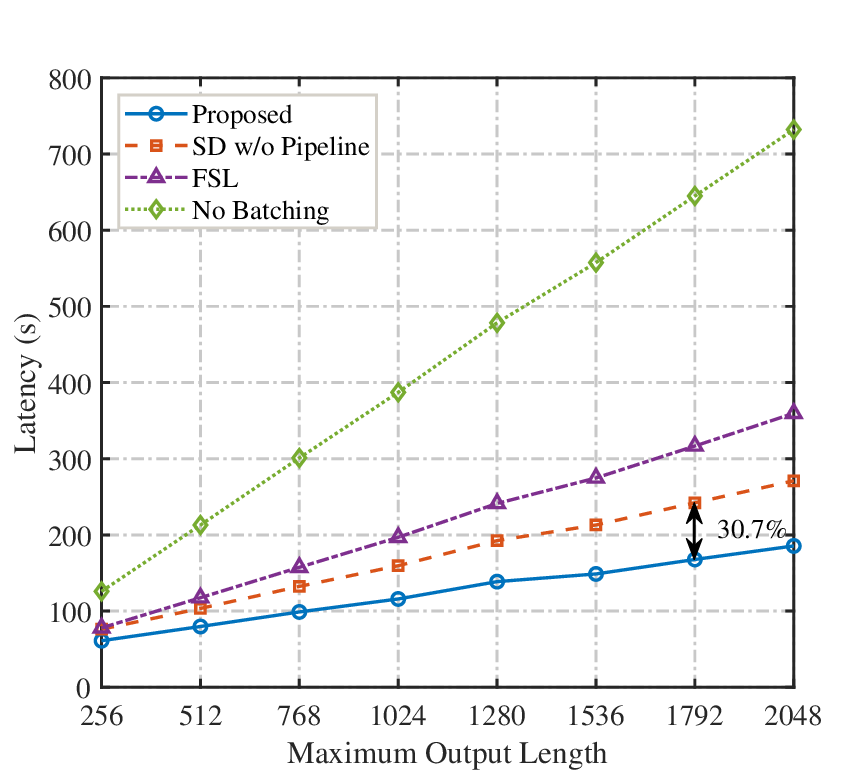} \\
			\small (a) & \small (b) & \small (c)
		\end{tabular}
		\caption{Performance comparison of the proposed method against baseline serving mechanisms under different parameters: (a) task number, (b) maximum input length, and (c) maximum output length.}
		\label{fig:serving_comparison}
		\vspace{-10pt} 
	\end{figure*}
	\newcommand{\figkserving}{\ref{fig:serving_comparison}(a)}
	\newcommand{\figiserving}{\ref{fig:serving_comparison}(b)}
	\newcommand{\figoserving}{\ref{fig:serving_comparison}(c)}
	Fig. \ref{fig:accep} presents the impacts of the acceptance rate (i.e., $\alpha$) on total latency for both the proposed serving mechanism and AD-based serving mechanism for three draft–verify model pairs.
	Note that different acceptance rates reflect the draft model's performance on various task types, with models typically exhibiting high acceptance rates for simple tasks (e.g., text completion) and low acceptance rates for complex tasks (e.g., mathematical reasoning).
	It is observed that compared to the AD-based serving mechanism, the proposed serving mechanism achieves consistently lower latency across all draft-verify model combinations and acceptance rates.
	In addition, as $\alpha$ grows, the latency reduction achieved by the proposed serving mechanism becomes more significant compared to ADS.
	The latent reason is that higher acceptance rates result in more draft tokens being accepted by the verify model in each decoding step, thereby generating more output tokens per decoding step.
	Consequently, fewer decoding steps are required to complete all tasks, reducing both inference and total latency.
	We also observe that, at the same acceptance rate, (LLaMA-1.1B, LLaMA-13B) exhibits higher latency than the other two model combinations.
	This is attributed to the larger verify model in (LLaMA-1.1B, LLaMA-13B), which contains more parameters and provides higher output accuracy.
	However, this larger verify model requires more FLOPs during execution, resulting in increased inference latency and consequently higher total latency.
	These results  indicate that the proposed serving mechanism not only consistently outperforms the AD-based serving system, but also effectively reduces inference latency across tasks of varying difficulty.

	Fig. \ref{fig:serving_comparison} shows the performance comparison of the proposed method against baseline serving mechanisms under different task configurations for the model pair (LLaMA-1.1B, LLaMA-7B) with an acceptance rate of $\alpha=0.8$.
	From Fig. \ref{fig:serving_comparison}(a), the latency of all serving mechanisms exhibits an approximately linear increase with the task number (i.e., $K$).
	This is because a larger number of tasks necessitates uploading more input sequences from users, thereby increasing communication latency.
	Additionally, more tasks increase the computational load, consequently increasing inference latency.
	It  is also seen that the proposed scheme consistently achieves lower latency than the three baseline methods across different task numbers.
	The underlying reasons for this performance gain are as follows:
	1) The proposed serving mechanism overlaps the draft and verification phases across batches at each decoding step.
	This effectively reduces the idle time of GPUs at both the SBS and MBS, thus achieving lower inference latency compared to the SD w/o pipeline scheme.
	Moreover, the proposed scheme incorporates adaptive speculation length control to effectively balance the latencies of both draft and verification phases, further minimizing GPU idle periods and achieving lower inference latency than the FSL scheme.
	2) Compared to the no-batching mechanism, the proposed serving mechanism dynamically partitions tasks into multiple batches and utilizes the parallel computing capacity of GPUs for batched processing.
	This enables processing multiple tasks with only minimal latency overhead compared to processing  tasks one by one, thereby reducing inference latency.
	
	Fig. \ref{fig:serving_comparison}(b) presents how the latency varies with the maximum input length (i.e., $I_{\rm{max}}$) across different serving mechanisms.
	One observation is that the latency of all serving mechanisms increases as $I_{\rm{max}}$ increases.
	Compared with the best baseline, i.e., SD w/o pipeline, the proposed serving mechanism achieves a 31.6\% latency reduction when $I_{\rm{max}}=512$.
	The reason is that increasing $I_{\rm{max}}$ leads to longer input sequences of tasks, which impose higher transmission loads and greater computational demands, thereby increasing both communication and inference latencies.
	Fig. \ref{fig:serving_comparison}(c) exhibits the latency versus maximum output length (i.e., $O_{\rm{max}}$) for different serving mechanisms.
	As expected, the latency of all schemes increases as the growth of $O_{\rm{max}}$.
	This is because the larger output length requires more decoding steps to process all tasks, thereby increasing the computational load and inference latency.
	Notably, as maximum output length increases, the performance advantage of the proposed mechanism becomes more pronounced compared to all benchmarks.
	Specifically,  the proposed serving mechanism reduces latency by about 30.7\% compared with SD w/o pipeline when maximum output length is 1792.
	These results verify that the proposed serving mechanism achieves low latency under varying computational and communication loads.

			\begin{figure*}[t]
		\centering
		\begin{tabular}{ccc}
			\includegraphics[width=0.32\textwidth]{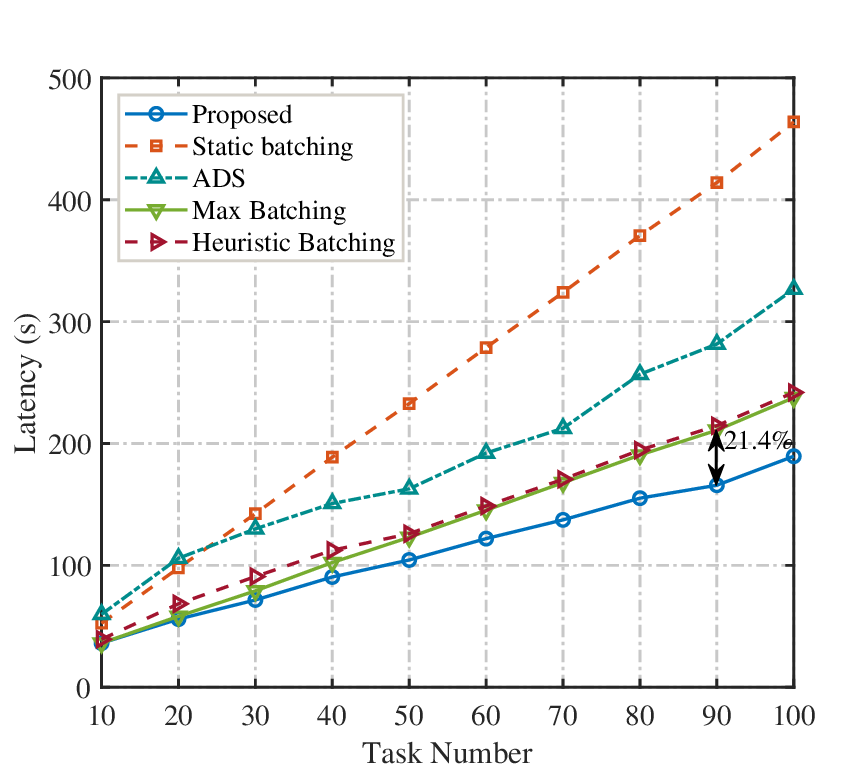} &
			\includegraphics[width=0.32\textwidth]{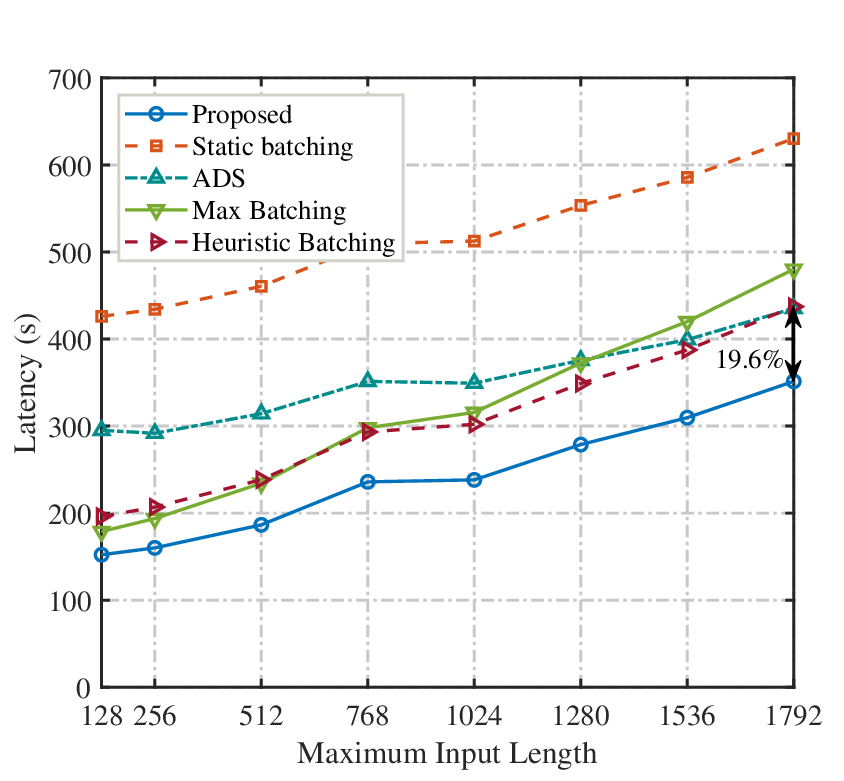} &
			\includegraphics[width=0.32\textwidth]{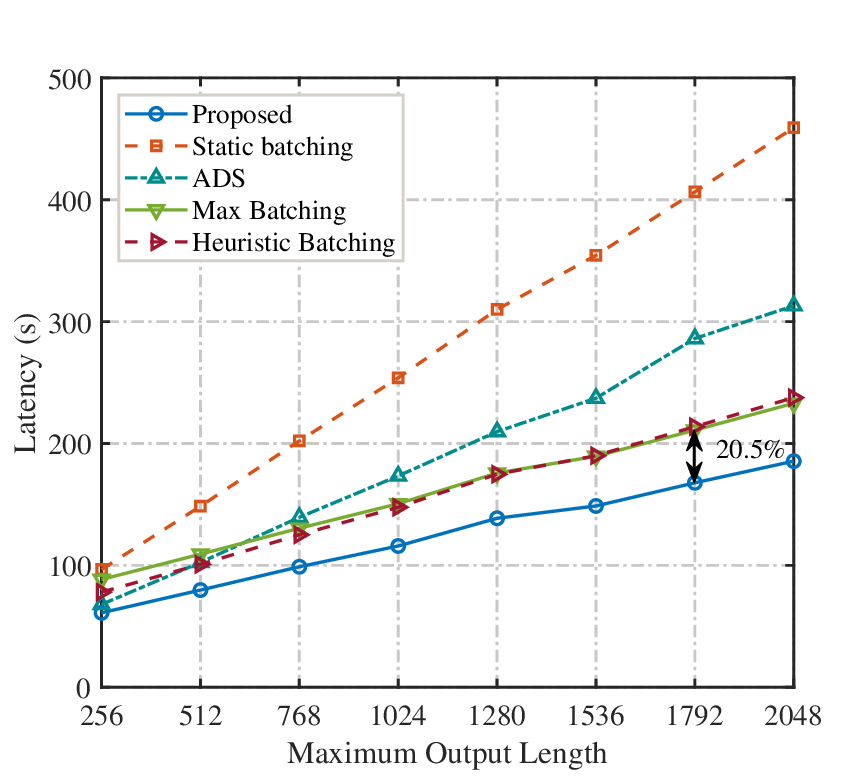} \\
			\small (a) & \small (b) & \small (c)
		\end{tabular}
		\caption{Performance comparison of the proposed batching scheme against baseline batching schemes under different parameters: (a) task number, (b) maximum input length, and (c) maximum output length.}
		\label{fig:batching_comparison}
										\vspace{-10pt} 
	\end{figure*}
	
	\newcommand{\figkbatch}{\ref{fig:batching_comparison}(a)}
	\newcommand{\figibatch}{\ref{fig:batching_comparison}(b)}
	\newcommand{\figobatch}{\ref{fig:batching_comparison}(c)}
		\vspace{-1.0em}
	\subsection{Performance of the Proposed Batching Algorithm}
	This subsection evaluates the proposed batching algorithm by comparing it with the following benchmarks.
	1) Static batching \cite{zhang2024beyond}: 
	The SBS and MBS implement the proposed LLM serving mechanism, where all tasks are assembled into batches with a small fixed batch size to prevent memory overflow.
	2) Max batching \cite{zhang2024edgeshard}: This scheme determines the maximum feasible batch size based on arriving tasks' input and output lengths, then uses this batch size to partition tasks in the proposed LLM serving mechanism.
	3) Heuristic batching \cite{chen2025spin}: The batch size is heuristically configured using the same strategy as in the AD-based serving mechanism. 
	Tasks are subsequently assembled into batches using this batch size in the proposed LLM serving mechanism.

	Fig. \ref{fig:batching_comparison} plots the impact of the batching schemes on the latency under different task configurations for model combination  (LLaMA-1.1B, LLaMA-7B) 
	under acceptance rate $\alpha=\textrm{0.8}$.
	From Fig. \ref{fig:batching_comparison}(a), we see that the latency of all batching schemes increases as the number of tasks grows, primarily due to the increased communication and computational loads.
	In addition, it is observed that the static batching scheme  exhibits the highest latency when the task number is large.
	This is because the static batching scheme groups all tasks into a small fixed batch size without considering either the current computational load or the characteristics of the tasks, which ultimately leads to significant inference latency.
	In contrast, the proposed batching scheme consistently outperforms all benchmarks across different task numbers.
	Specifically, when the task number reaches 90, the proposed batching scheme achieves up to a 21.4\% reduction in latency compared with the max batching scheme.
	The performance gain comes from the proposed scheme’s ability to efficiently group tasks with  similar input lengths into the same batch, which reduces padding and avoids inefficient computational overhead, thereby further lowering inference latency. 
	In addition, the synergy between the proposed batching scheme and adaptive speculation length control enables efficient coordination of the draft phases of the current batch and the verification phase of the next batch. 
	This reduces GPU idle time at both the SBS and MBS and thus decreases the overall serving latency.

	Fig. \ref{fig:batching_comparison}(b) and Fig. \ref{fig:batching_comparison}(c)
	present the latency versus the maximum input length and output length, respectively.
	As before, the proposed scheme outperforms all benchmarks under different maximum input lengths and output lengths. 
	Specifically, compared with the best benchmark (i.e., heuristic batching scheme), 
	the proposed scheme reduces latency by up to 19.6\% when the maximum input length is 1792, and by 20.5\% when the maximum output length is 1024.
	Moreover, it is observed that the proposed scheme also achieves lower latency than the AD-based serving mechanism under different settings of task number, maximum input length, and maximum output length.
	These results verify that the proposed batching scheme efficiently adapts to system load and task characteristics, i.e., input and output lengths, to achieve lower latency.
	
	\begin{figure*}[t]
		\centering  
		\begin{tabular}{ccc}
			\includegraphics[width=0.32\textwidth]{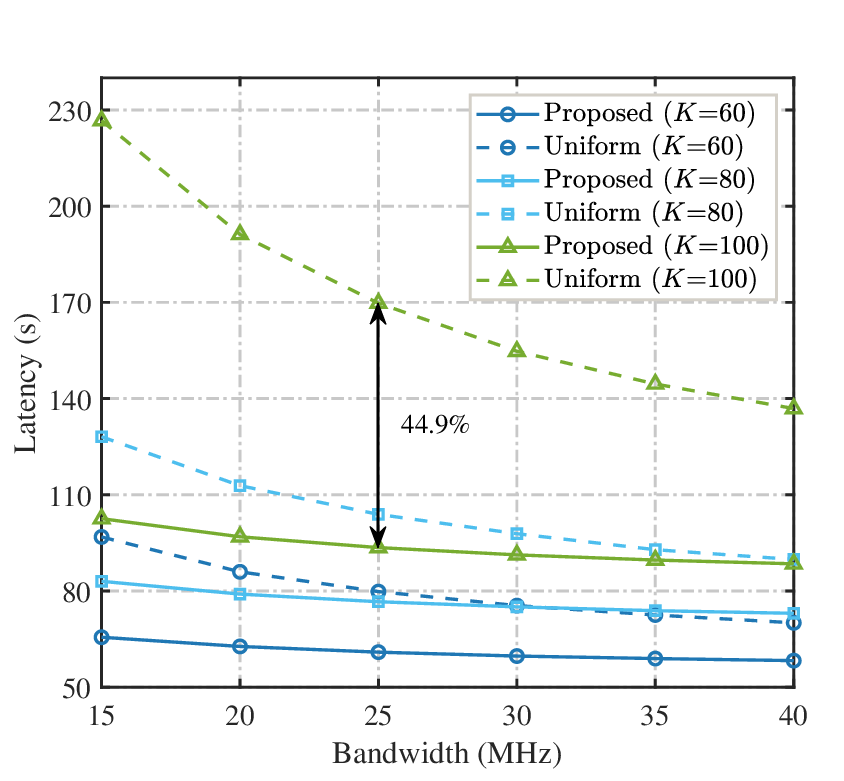} &
			\includegraphics[width=0.32\textwidth]{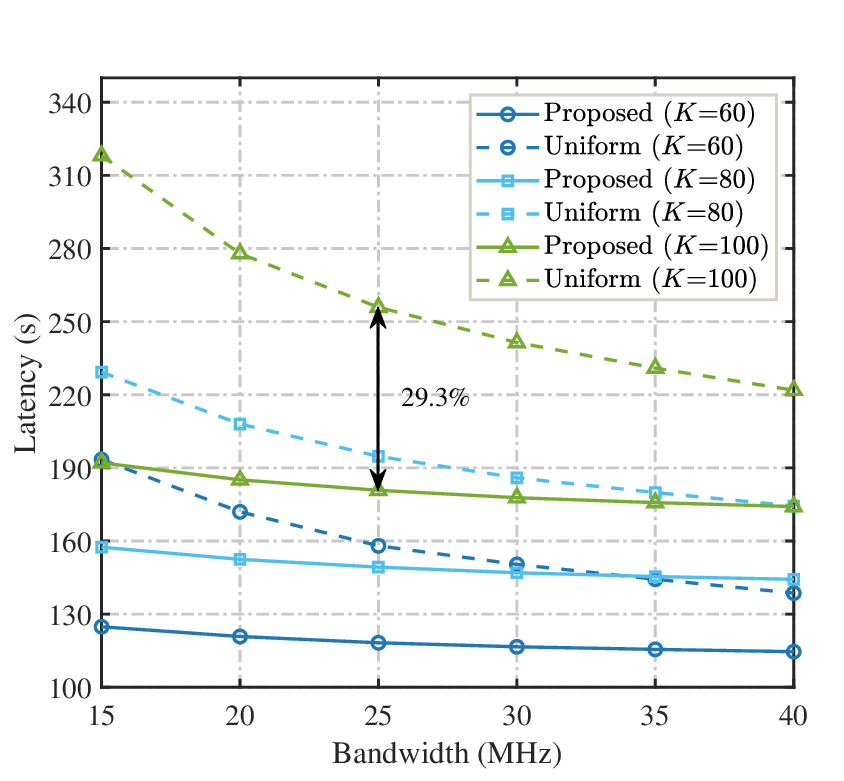} &
			\includegraphics[width=0.32\textwidth]{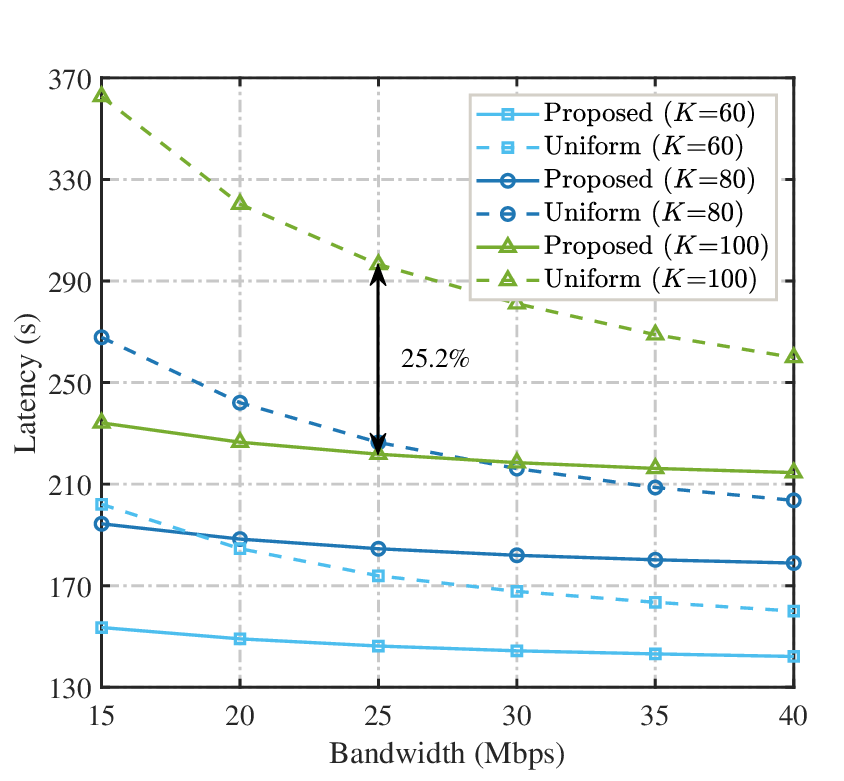} \\
			\small (a) & \small (b) & \small (c)
		\end{tabular}
		\caption{Performance comparison of the proposed communication resource allocation scheme against uniform schemes under different bandwidth: (a) model combination (LLaMA-68M, LLaMA-7B), (b) model combination (LLaMA-1.1B, LLaMA-7B), and (c) model combination (LLaMA-1.1B, LLaMA-13B).}
		\label{fig:bandw}
		\vspace{-10pt} 
	\end{figure*}
		\vspace{-0.8em}
	\subsection{Performance of the Proposed Communication Resource Allocation Algorithm} \label{sub:sim_com}
	In this subsection, we evaluate the proposed communication resource allocation policy by comparison with a uniform resource allocation scheme. 
	In the uniform scheme, the available uplink bandwidth is evenly divided among the uplink transmissions of all tasks.
	
	Fig.~\ref{fig:bandw} illustrates how latency varies with the available bandwidth (i.e., $B_{\rm w}$) for different task numbers (i.e., $K$) across three model combinations at $\alpha=0.8$.
	As shown in Fig.~\ref{fig:bandw}, for each model combination, the latency of both resource allocation schemes decreases as the bandwidth increases.
	This is attributed to the fact that larger  bandwidth enables higher wireless transmission rates for uploading tasks' input sequences, thereby reducing communication and total latency.
	It is also observed that the latency of both schemes increases as the number of tasks grows.
	The reason is that a larger number of tasks results in less wireless bandwidth allocated per task, leading to lower transmission rates and consequently higher communication latency.
	Moreover, the proposed resource allocation scheme consistently outperforms the uniform allocation scheme across different bandwidths and task numbers.
	Specifically, 
	Fig.~\ref{fig:bandw}(a) shows that the proposed scheme reduces latency by 44.9\% compared with the uniform allocation scheme when the number of tasks is $K=100$, the wireless bandwidth is $B_{\rm w}=25$ MHz, and the model combination is (LLaMA-68M, LLaMA-7B).
	This performance gain arises because the proposed scheme adaptively allocates wireless bandwidth based on users’ channel conditions, ensuring that tasks within the same batch arrive at the SBS simultaneously and thereby reducing communication latency.
	The results validate that the proposed scheme effectively adapts to varying numbers of tasks and achieves low latency even when bandwidth is small.
	
	The experiments with (LLaMA-1.1B, LLaMA-7B) and (LLaMA-1.1B, LLaMA-13B), shown in Fig.~\ref{fig:bandw}(b) and Fig.~\ref{fig:bandw}(c), yield similar conclusions.
	Specifically, when the system contains $K=100$ tasks and the bandwidth is $B_{\rm w}=25$ MHz, the proposed scheme reduces latency by up to 29.3\% and 25.2\% for (LLaMA-1.1B, LLaMA-7B) and (LLaMA-1.1B, LLaMA-13B), respectively, compared with the uniform scheme.
	Furthermore, we see that among the three model combinations, (LLaMA-68M, LLaMA-7B) exhibits the lowest latency. 
	This can be attributed to the relatively small model dimensions of both LLaMA-68M and LLaMA-7B, i.e., $h_1^{\rm d}$ and $h_1^{\rm v}$. 
	According to \eqref{eq:ul_latency_k}, smaller model hidden dimensions reduce the embedding size per input token, i.e., $\lambda=16(h_1^{\rm d}+h_1^{\rm v})$, thereby decreasing the total uplink transmission data size and communication latency. 
	In addition, the small model dimensions lead to low computational cost, further decreasing inference latency.
	Consequently, (LLaMA-68M, LLaMA-7B) achieves the minimum latency among the three combinations.

	\section{Conclusion}\label{sec:conclusion}
	In this work, we developed a novel SD-based LLM serving framework that distributes draft and verify models on a heterogeneous edge network to deliver low-latency LLM inference services.
	Then, we proposed a pipeline parallel-enabled serving mechanism that overlaps the draft and verification phases of different inference tasks, thereby reducing GPU idle time at both the SBS and MBS and thus reduce serving latency.
	Based on this framework, we developed a latency model to effectively characterize the serving latency of inference tasks.
	Then, we developed effective joint optimization policies for wireless communication resource allocation, batching, and speculation length control to minimize the total serving latency of inference tasks under limited communication and memory resources.
	Experimental results verified that the proposed LLM serving system efficiently reduces latency compared with AD-based LLM serving systems.
	Additionally, experimental results demonstrated that the proposed optimization policies achieved lower latency than baseline methods.

	\bibliographystyle{IEEEtran}
	\bibliography{IEEEabrv,cited}

\end{document}